\documentclass[superscriptaddress,twocolumn,amssymb,aps]{revtex4}

\usepackage{graphicx,dcolumn,bm,amsmath,color,bm}

\newcommand{\eq}[1]{Eq.~(\ref{#1})}
\newcommand{\fig}[1]{Fig.~\ref{#1}}
\newcommand{\avg}[1]{ {\langle #1 \rangle} }

\newcommand{\eeq}{ \end{equation} }
\newcommand{\beq}{ \begin{equation} }

\newcommand{\bhu}{ {\bf \hat{u}} }
\newcommand{\bhe}{ {\bf \hat{e}} }

\newcommand{\bfr}{ {\bf r} }

\newcommand{\bn}{ {\bf \hat{n}} }
\newcommand{\bhatm}{ {\bf \hat{m}} }
\newcommand{\bx}{ {\bf \hat{x}} }
\newcommand{\by}{ {\bf \hat{y}} }
\newcommand{\bz}{ {\bf \hat{z}} }
\newcommand{\bv}{ {\bf \hat{v}} }

\newcommand{\btau}{ {\bf \hat{\tau}} }
\newcommand{\bchi}{ {\bf \hat{\chi}} }

\newcommand{\bQ}{ {\bf Q}^{\rm (m)} }
\newcommand{\Sm}{ S_{\rm m} }
\newcommand{\Seq}{ S_{\rm{eq}}^{\rm (m)} }
\newcommand{\pdQm}[2]{ {\frac{\partial \bQ_{#1}}{\partial x_{#2}}} }
\newcommand{\pdn}[2]{ {\frac{\partial \bn_{#1}}{\partial x_{#2}}} }
\newcommand{\red}[1]{ { \color{red} #1 } }

\newcommand{\ul}{\rm{\mu l}}
\newcommand{\um}{\rm{\mu m}}
\newcommand{\Dc}{D_{\rm{c}}}
\newcommand{\Lc}{L_{\rm{c}}}
\newcommand{\hide}[1]{}

\begin{document}

\title{Unavoidable emergent biaxiality in chiral molecular-colloidal hybrid liquid crystals}

\author{J.-S. Wu}
\affiliation{Department of Physics and Chemical Physics Program, University of Colorado, Boulder, CO, USA}
\author{M. Torres L\'{a}zaro}
\affiliation{Laboratoire de Physique des Solides - UMR 8502, Universit\'e Paris-Saclay  \& CNRS,  91405 Orsay, France}
\author{S. Ghosh}
\affiliation{Department of Physics and Chemical Physics Program, University of Colorado, Boulder, CO, USA}
\author{H. Mundoor}
\affiliation{Department of Physics and Chemical Physics Program, University of Colorado, Boulder, CO, USA}
\author{H. H. Wensink}
\affiliation{Laboratoire de Physique des Solides - UMR 8502, Universit\'e Paris-Saclay  \& CNRS,  91405 Orsay, France}
\author{I. I. Smalyukh}
\affiliation{Department of Physics and Chemical Physics Program, University of Colorado, Boulder, CO, USA}
\affiliation{International Institute for Sustainability with Knotted Chiral Meta Matter, Hiroshima University, Higashihiroshima, Japan}
\affiliation{Department of Electrical, Computer, and Energy Engineering, Materials Science and Engineering Program and Soft Materials Research Center, University of Colorado, Boulder, CO, USA}
\affiliation{Renewable and Sustainable Energy Institute, National Renewable Energy Laboratory and University of Colorado, Boulder, CO 80309, USA}

\email{ivan.smalyukh@colorado.edu}

\date{\today}

\begin{abstract}
Chiral nematic or cholesteric liquid crystals (LCs) are chiral mesophases with long-ranged orientational order featuring a quasi-layered periodicity imparted by a helical director configuration but lacking positional order. Doping molecular cholesteric LCs with thin colloidal rods with a large length-to-width ratio or disks with a large diameter-to-thickness ratio adds another level of complexity to the system because of the interplay between weak surface anchoring boundary conditions and bulk-based elastic distortions around the particle-LC interface. By using colloidal disks and rods with different geometric shapes and boundary conditions, we demonstrate that these anisotropic colloidal inclusions exhibit biaxial orientational probability distributions, where they have tendencies to orient with the long rod axes and disk normals perpendicular to the helix axis, thus imparting strong local biaxiality on the hybrid cholesteric LC structure. Unlike the situation in non-chiral hybrid molecular-colloidal LCs, where biaxial order emerges only at modest to high volume fractions of the anisotropic colloidal particles, above a uniaxial-biaxial transition concentration, the orientational probability distribution of colloidal inclusions immersed in chiral nematic hosts are unavoidably biaxial even at vanishingly low particle volume fractions. In addition, the colloidal inclusions induce local biaxiality in the molecular orientational order of the LC host medium, which enhances the weak biaxiality of the LC in a chiral nematic phase coming from the symmetry breaking caused by the presence of the helical axis. With the help of analytical modeling and computer simulations based on minimizing the Landau de Gennes free energy of the host LC around the colloidal inclusions, we explain our experimental findings and conclude that the biaxial order of chiral molecular-colloidal LCs is strongly enhanced as compared to both achiral molecular-colloidal LCs and molecular cholesteric LCs and is rather unavoidable.  
\end{abstract}

\maketitle
\section{Introduction}
Since the experimental discovery of chiral nematic liquid crystals (LCs) over 150 years ago \cite{planer1861notiz,reinitzer1888beitrage}, LC mesophases featuring chirality and long-range orientational order have been the focus of many research studies. The fundamental studies of geometry and topology of chiral nematic LCs as model systems provide extensive insights into physics principles associated with experimentally less accessible systems like particle physics or cosmology \cite{wu2022hopfions,ackerman2014two,tai2020surface,volovik2001superfluid,bowick1994cosmological,meng2023topological,smalyukh2020knots,smalyukh2018liquid,kamada2022chiral,chaikin1995principles}, in addition to their technological application in electro-optics and displays. On the other hand, biaxial nematic mesophases have been highly sought-after in soft matter systems since their first theoretical consideration in 1970 \cite{freiser1970ordered}. However, even in a soft-matter system with strongly biaxial building blocks such as brick-shaped molecules, biaxiality was experimentally elusive and hard to unambiguously demonstrate in equilibrium states. Recent reports of the experimental discovery of biaxial nematic order include observations in micellar and molecular LCs formed by amphiphilic and bent-core molecules, respectively \cite{yu1980observation,tschierske2010biaxial}, and also colloidal dispersion of highly anisotropic particles immersed in molecular LC hosts, so-called hybrid molecular-colloidal nematics \cite{liu2016,mundoor2021,mundoor2018}.
The interplay between chirality and biaxiality in orientational order has been intensively studied for LC systems \cite{priest1974biaxial,kroin1989chirality,bunning1986effect,harris1997microscopic,dussi2016entropy,dhakal2011chirality,longa1994biaxiality,canevari2015biaxiality,harris1999molecular,lubensky1998chirality}. It has been concluded that cholesteric twisted alignment and biaxial order of LC molecules amplify each other and that a chiral twist configuration cannot be observed without building blocks featuring a certain degree of biaxiality in their orientational distributions at the molecular level. However, for purely molecular systems, the chirality-enhanced biaxiality of the molecular distribution was predicted and experimentally found to be rather weak \cite{priest1974biaxial,kroin1989chirality,bunning1986effect,harris1997microscopic,dussi2016entropy,dhakal2011chirality,longa1994biaxiality,canevari2015biaxiality}, scaling as $\ (qL_{\rm m})^2$ according to the prediction by Priest and Lubensky for single-component molecular LCs \cite{priest1974biaxial} (here $q=2\pi/p$, $p$ is the helical pitch of the chiral nematic and $L_m$ the molecular length). To date, to the best of our knowledge, there are no experimental or theoretical considerations of how the biaxiality of the orientational distribution of anisotropic colloidal particles could interplay with the chirality of the nematic host in hybrid molecular-colloidal LC systems.  

\begin{figure}
	\includegraphics[width=0.9\columnwidth]{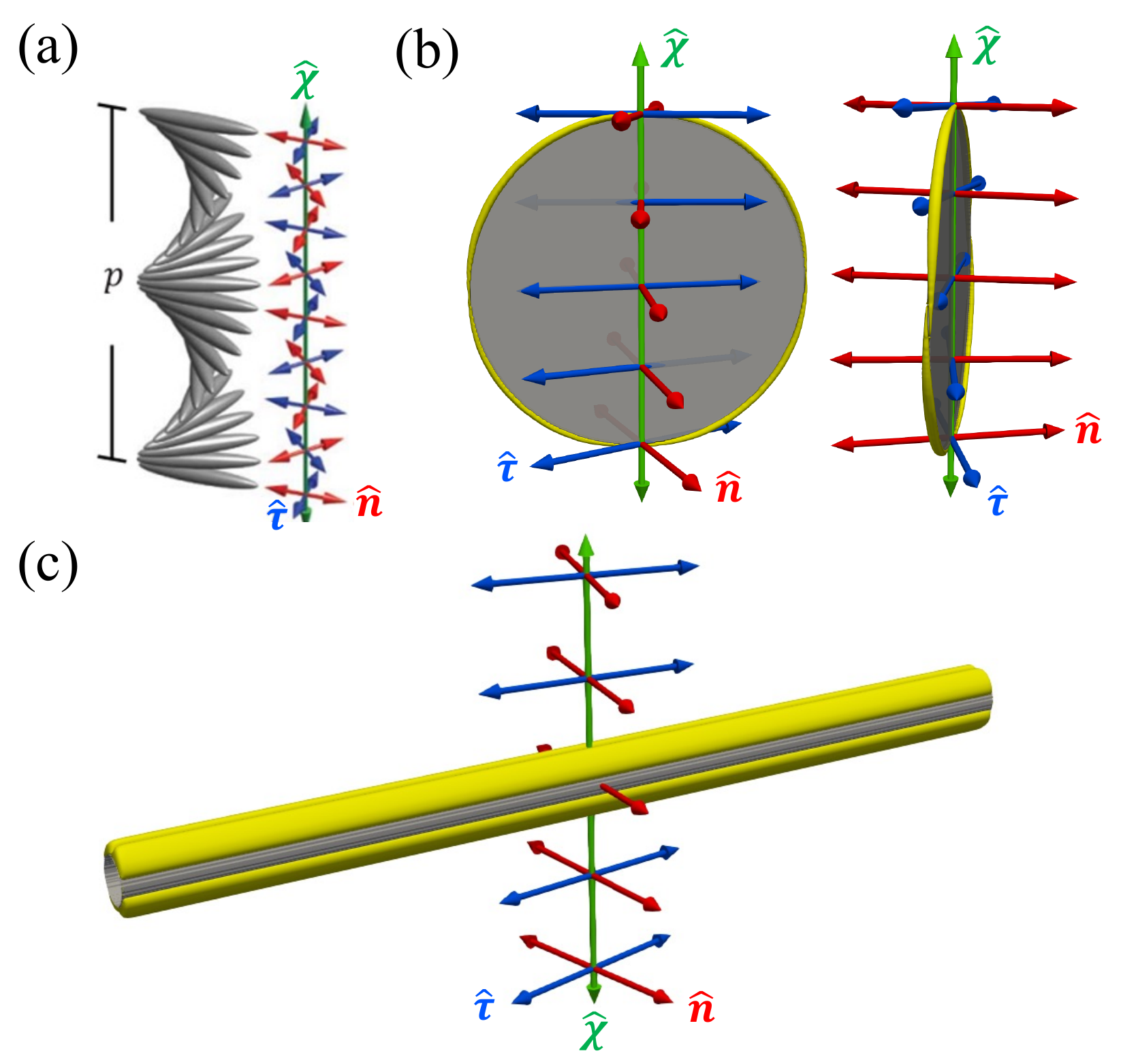}
	\caption{(a) Helical structure of a chiral LC with helical pitch length $p$, with gray ellipsoids representing LC molecules and colored axes depicting the orthogonal molecular frame: LC director $\bn$ (red), helical axis $\bchi$ (green), and the third axis $\btau$ (blue). (b)-(c) Visualizations of a colloidal disk (b) and rod (c) immersed in a chiral LC at their equilibrium orientations. Colloids are colored in gray, and the yellow contours mark a director deviation of 0.67° (b) and 0.3° (c), respectively, of the numerically-relaxed LC structures from the ideal helical state indicated by the colored double arrows. For all simulations the anchoring at the colloid surface is homeotropic with strength $W_{0} = 10^{-4} {\rm Jm^{-2}}$.}
	\label{chiral_schematic}
\end{figure}

In this work, we demonstrate an unavoidably biaxial orientation probability distribution for uniaxial colloidal particles dispersed in a weakly chiral molecular host, which is rather unexpected. We report strongly enhanced biaxial order in the orientational distribution probability for colloidal rods scaling as $(qL_{\rm c})^2$, where the length of the colloidal particles $L_{\rm c}$ is in the micron range, more than 3 orders of magnitude larger than that of LC molecules, albeit still an order of magnitude or so smaller than the pitch $p$. The geometry of the cholesteric LC is described by three non-polar, orthogonal director fields [\fig{chiral_schematic}]: molecular director field $\bn=-\bn$ representing the local average molecular alignment, the helical axis field $\bchi=-\bchi$ along which $\bn$ rotates, and the third field $\btau=\pm \bn \times \bchi$ \cite{kleman1989defects,lavrentovich2001cholesteric}.
Here, the helicoidal configuration with a mutually perpendicular molecular frame $(\bn,\bchi,\btau)$ and helical pitch $p$ is hardly perturbed by the introduction of thin colloidal disks or rods in view of their low concentration and weak surface anchoring boundary conditions.
The colloidal particles are uniaxial on their own with high aspect ratios, and their orientations are well-controlled by pre-designed boundary conditions. The biaxiality of the colloidal orientational distribution is found to exceed values known for biaxiality on the molecular scale in cholesteric LCs and in nematic molecular colloidal-molecular hybrid systems, despite the low colloidal concentration and weak chirality of the molecular host. 
In cases where the colloids align along the $\btau$ axis, the biaxiality is found to be more pronounced than when they are aligned along the molecular director $\bn$. Furthermore, the molecular biaxiality of the LC host medium is further boosted by surface anchoring-induced distortions at the edges of the colloidal particles. 
In contrast to the orientational distribution of colloidal inclusions in nematic hybrid molecular-colloidal LCs, in which biaxial order emerges only at modest to high volume fractions of anisotropic colloidal particles \cite{mundoor2021}, the orientational probability distribution of colloidal inclusions in chiral nematic hosts is unavoidably biaxial even at very low colloidal volume fractions.
In turn, colloidal inclusions impart local biaxiality onto the molecular orientational distributions of the LC host medium, which enhances the weak biaxial order of the LC in a chiral nematic phase due to the symmetry breaking caused by the presence of the helical axis. With the help of computer simulations based on minimizing the Landau de Gennes free energy of the LC host around the colloidal inclusions [\fig{numerical_visualization}], we explain our experimental findings and conclude that the biaxial order of chiral molecular-colloidal LCs is enhanced compared to that of both nematic molecular-colloidal and cholesteric molecular counterparts. We demonstrate that the interplay between chirality and biaxiality in hybrid molecular-colloidal LCs is stronger than that in purely molecular or colloidal systems and that the biaxial symmetry of the orientational distributions of the anisotropic colloids is a universal feature. Finally, we discuss how our findings may allow for expanding the use of chiral molecular-colloidal LCs as model systems in studies of nonabelian defect lines and topological solitons hosted by states of matter with high-dimensional order parameter spaces. 

\section{Methods and Techniques}
\subsection{Synthesis of colloidal disks and rods}

Disk or rod-shaped $ \beta - \rm{NaYF_4}$:Yb/Er particles are synthesized following the hydrothermal synthesis methods described in detail elsewhere \cite{mundoor2021,mundoor2018,mundoor2019electrostatically,wang2007controlled,yang2012one}. 
Precursors and solvents used for the synthesis of colloidal particles are of analytical grade and used without additional purifications, and they are bought from Sigma Aldrich if not specified otherwise.
 \hide{ Ytterbium nitrate hexahydrate ($\rm{Yb(NO_3)_3 \cdot 6H_2O}$), yttrium nitrate hexahydrate ($\rm{Y(NO_3)_3 \cdot 6H_2O}$), erbium nitrate pentahydrate ($\rm{Er(NO_3)_3 \cdot 5H_2O}$), sodium fluoride, oxalic acid, and oleic acid are purchased from Sigma Aldrich, and sodium hydroxide is purchased from Alfa Aesar.}  
To synthesize nanodisks, 0.7 g of sodium hydroxide (purchased from Alfa Aesar) is dissolved in 10 ml of deionized water and then added with 5 ml of oxalic acid solution (2g, 19.2 mmol) at room temperature to obtain a transparent solution. 
Under vigorous stirring, we then add 5 ml of sodium fluoride solution (202 mg, 4.8 mmol) to the mixture. After 15 minutes of stirring, 1.1 ml of $\rm{Y(NO_3)_3}$ (0.88 mmol), 0.35 ml of $\rm{Yb(NO_3)_3}$ and 0.05 ml of $\rm{Er(NO_3)_3}$ are added into the mixture while the stirring continues for another 20 minutes at room temperature.
Subsequently, the solution is transferred to a 40-ml Teflon chamber (Col-Int. Tech.) and heated to and kept at 200 °C for 12 hours (h). 
The mixture is then cooled down naturally to room temperature, and the particles precipitated at the bottom are collected by centrifugation, rinsed with deionized water multiple times, and finally dispersed in 10 ml of deionized water.
Colloidal rods are prepared using a similar protocol: 1.2 g of NaOH is dissolved in 5 ml of deionized water and mixed with 7 ml of ethanol and 20 ml of oleic acid under stirring, followed by adding 8ml of NaF (1 M), 950 $\ul$ of $\rm{Y(NO_3)_3}$ (0.5 M), 225 $\ul$ of $\rm{Yb(NO_3)_3}$ (0.2 M), and 50 $\ul$ of $\rm{Er(NO_3)_3}$ (0.2 M) and stirring for 20 minutes. The obtained white viscous mixture is transferred into a 50 ml Teflon chamber, kept at 190°C for 24 h, and then cooled down to room temperature. The particles deposited at the bottom of the Teflon chamber are collected and washed with ethanol and deionized water multiple times and finally dispersed in cyclohexane.

In some cases, silica microrods synthesized following an emulsion-templated wet-chemical approach \cite{kuijk2011synthesis} are also used. To synthesize them, 1 gm of polyvinylpyrrolidone (PVP, molecular weight 40000) is dissolved in 10 ml of 1-pentanol, followed by the addition of 950 $\ul$ of absolute ethanol (Decon labs), 280 $\ul$ of deionized water, 100 $\ul$ of sodium citrate solution (0.18 M), and 130 $\ul$ of ammonia solution (28\%). The bottle is shaken vigorously using a vortex mixer after each addition.
Then, 100 $\ul$ of tetraethyl orthosilicate (TEOS, 98\%) is added under agitation. The bottle is incubated at 25 °C for the next 8 h. The solution becomes milky white after the reaction, and it is centrifuged at 6000 revolutions per minute (RPM) for 10 minutes to separate the as-synthesized rods. The precipitated rods are then washed two times with water followed by another two rounds of washing with ethanol at 3000 RPM for 5 minutes. Finally, to improve the monodispersity and to remove other lightweight impurities, the rods are centrifuged at 500 RPM for 30 minutes and dispersed in ethanol, with the procedure repeated two more times.

\subsection{Surface functionalization of the colloids}
Homeotropic surface anchoring boundary conditions for the director and 5CB (pentylcyanobiphenyl or 4-cyano-4'-pentylbiphenyl) molecules on the $\beta$-$\rm{NaYF_4}$:Yb/Er disk surfaces is controlled through surface-functionalization with a thin layer of silica and polyethylene glycol.
First, 5 ml of hydrogen peroxide ($\rm{H_2O_2}$) is added to 1 ml of colloidal disk dispersion in deionized water. Then, under vigorous mechanical agitation, 100 $\ul$ of nitric acid is added drop by drop into the solution.
\hide{ During this reaction, the disks acquired positive surface charges, while oxalic acid molecules attached to the surfaces of the particles were oxidized. }
After 12 h of agitation, disks are separated from the liquid by centrifugation and transferred into 1 ml of ethanol. 
The colloidal dispersion is then mixed with 75 mg of polyvinyl pyrrolidone (molecular weight 40,000) in 4 ml of ethanol and kept under continuous mechanical agitation for another 24 h. 
The particles are collected and redispersed in 5 ml of ethanol, before the addition of 200 $\ul$ of ammonia solution and 6 $\ul$ of tetraethyl orthosilicate under mechanical agitation that lasts 12 h.
Disks are collected, washed with ethanol and deionized water, and redispersed in 4 ml of ethanol. 
The pH value of the mixture is adjusted to 12 by adding ammonia solution (28\% in water). Then, under mechanical agitation at 35°C, we add 35 mg of silane-terminated polyethylene glycol (molecular weight 5,000, dissolved in 1 ml of ethanol at 50°C) to the solution.
After another 12 h of agitation, the surface-functionalized disks are again collected, washed with ethanol and water, and dispersed in 1 ml of ethanol.

As for the hydrothermal-synthesized rods, the surface chemical treatment not only provides the desired anchoring preference but also controls the cylinder aspect ratio.
For this, 4 ml of the nanorod dispersion is added with 200 $\ul$ of HCl in 2 ml of water and kept stirred overnight. 
The nanorods are then transferred from organic to aqueous phases. The nanorods are collected by centrifugation, washed with deionized water and ethanol three times, dispersed in deionized water, and then finally re-dispersed in ethanol.
The process of etching with acid and redispersion is repeated two more times, with HCl treatments of 12 hours and 3 hours, respectively. 
The aspect ratio of the nanorods is increased during acid treatment to an average value of $L_{\rm c}/D_{\rm c} \approx 60$.

Similarly, the emulsion-templated rods are slowly etched in a mild basic condition \cite{hagemans2016synthesis} with 0.5 mM NaOH for 24 h, followed by drying at 80°C for another 4 h. After this, the functionalization of silica rods is done by adding 100 $\ul$ of perfluorooctyltriethoxysilane (TCI America) to 0.9 mL ethanol dispersion of the silica rods. The mixture is kept at room temperature for 3 h before being washed and redispersed three times in ethanol.
After vacuum-drying inside a desiccator and heating at 60 °C for 1 h, the microrods are immersed in a perfluorocarbon liquid (Fluorinert FC-70, Alfa Aesar) and kept at 60 °C for 1 h before being cooled down to the room temperature and redispersed into ethanol for storage.
The fusion of perfluorocarbon oil onto the perfluorosilane functionalized rods results in a fully covered and stable slippery surface layer, giving desired boundary conditions.

\subsection{Colloidal particle dispersion in chiral molecular LC} 
A small amount of left-handed chiral dopant cholesterol pelargonate \hide{(Sigma Aldrich)}is added into molecular 5CB (Frinton Labs and Chengzhi Yonghua Display Materials Co. Ltd). To obtain the equilibrium pitch $p$ of the molecular chiral mixture, the weight fraction of the used chiral additive is roughly estimated by $c_d=\frac{1}{6.25p}$. The actual pitch is later revealed using optical microscopy by observing the periodicity of defect lines in Gradjean-Cano wedge cells \cite{smalyukh2002three}. The surface-functionalized particles are then dispersed into such molecular chiral LC.
In a typical experiment, 20 $\ul$ of colloidal dispersion in ethanol is mixed with 20 $\ul$ of the molecular LC. The mixture is then heated to 75°C and kept for 2 h to completely evaporate the organic solvent. A well-dispersed colloidal-molecular hybrid LC is usually obtained after quenching back to room temperature under mechanical agitation \cite{liu2014electrically,mundoor2015mesostructured,evans2011alignment}. Additional centrifugation can be carried out to remove the particle aggregation formed during the isotropic to chiral nematic phase transition of the molecular LC.

Hybrid LCs containing the colloidal dispersion are infiltrated into glass cells with gap thickness typically chosen to be between $p/2$ and 10$p$, which is experimentally set using Mylar films or silica spheres. To achieve unidirectional planar boundary conditions for 5CB molecules, cell substrates are coated with 1wt.\% aqueous polyvinyl alcohol \hide{(Sigma Aldrich) }and rubbed unidirectionally. Typically, the geometry and planar boundary conditions of the cell give a sample with its helical axis $\bchi$ perpendicular to the glass substrate and with the helical twist of the cholesteric host LC in compliance with the designed boundary conditions at the confining glass surfaces.

\subsection{Optical microscopy and characterization of colloidal orientations}
We use different optical microscopy methods to visualize the colloidal orientations inside the hybrid LC, among which are three-photon excitation fluorescence polarizing microscopy (3PEFPM), photon-upconverting confocal microscopy, polarizing optical microscopy and phase contrast microscopy.
Using 3PEFPM, optical imaging of director structures of the molecular host medium is performed using a multimodal 3-dimensional (3D) nonlinear imaging system built around a confocal system FV300 (Olympus) and an inverted microscope (Olympus IX-81) \cite{mundoor2015mesostructured,lee2010multimodal}.  The 3D imaging of the $\beta$-$\rm{NaYF_4}$:Yb/Er particles designed to exhibit upconversion luminescence is performed with the same setup when the colloidal dispersions are excited with a laser light at 980 nm; this photon-upconversion-based imaging of colloidal particles minimizes the background signal from the molecular LC, making such a technique ideal for our study.
A $100\times$ objective (Olympus UPlanFL, numerical aperture 1.4) and a 980-nm pulsed output from a Ti:Sapphire oscillator (80 MHz, Coherent, Chameleon ultra) are utilized, along with a set of Galvano mirrors on the optical path to achieve sufficient positional accuracy while scanning the sample horizontally. In addition, the vertical re-positioning is achieved by a stepper motor on which the objective could be adjusted to focus at the desired sample depth, enabling 3D scanning with high accuracy.
Luminescence signals are epi-collected using the same objective before being sent through a pinhole and detected by a photomultiplier tube. The data obtained from several scanning planes are combined into a 3D tiff image to be analyzed at a later time. 
The phase contrast images are taken using a $60\times$ objective (Olympus UPlanFL N, variable numerical aperture 0.65-1.25), mounted on another microscope system (Olympus IX-83), at various vertical positions controlled by a motorized sample stage.

The colloidal orientations, representing the normal direction of disks or the long axis of rods, are analyzed on the basis of two-dimensional (2D) slices of a 3D sample using ImageJ software (freeware from the National Institute of Health, \cite{schneider2012nih}), with the error in measured colloidal angles is about $\pm 1^{\circ}$. The ensuing statistical data are transferred to Matlab software for visualization, as well as for further analysis. The color thresholds of the images are carefully adjusted to avoid the interference of colloids out of focus. From the 3D stacks of images, the slice plane perpendicular to the helical axis $\bchi$ gives the azimuthal orientational distribution ($\varphi$\hide{$\delta$ for colloids aligning along $\bn$ and $\gamma=\delta+90 ^{\circ}$ for rods along $\btau$}), whereas the vertical slice plane reveals the polar distribution ($\theta$\hide{$\zeta$ for colloids along $\bn$ and $\eta$ for those along $\btau$}). Since the colloidal orientations are highly confined, as shown by the high value of uniaxial order $S_{\rm{cc}}$, we assume that the two distributions are independent so that the overall distribution can be written in factorized form $f(\varphi,\theta)=f(\varphi)f(\theta)$. For the same reason, we ignore the effect of the projection from the 3D volume to the slice planes. 
After the analysis of particles by ImageJ, average azimuthal colloidal orientations are calculated for the data obtained in each $\bn-\btau$ slice plane and plotted against the sample depth ($z$) position of the cross-sectional plane, revealing the helical twist of the colloidal axes. The corresponding helical pitch $p$ of each 3D volume is subsequently calculated from the slope of the linear dependence of the azimuthal angle on the vertical position ($d\varphi/dz=q=360^{\circ}/p$) and is in agreement with the initially designed value mentioned above, confirming the undisturbed molecular helical pitch at low colloidal concentrations.
Finally, the colloidal orientation distribution is visualized in the molecular director coordinate system frame. 
The azimuthal angle in the molecular coordinate is calculated by subtracting the molecular twist from the measured colloidal orientations, $\delta=\varphi-qz$ representing the fluctuation of colloidal orientation around that of a perfect helix. \hide{, while the data for the polar angle are not further transformed.}
The non-orientable property of the colloidal axis $\bhu=-\bhu$ enabled us to express the fluctuation angles, $\delta=\delta+\pi$, for example, within a [-90°,90°] range. 
Histograms of angular probability distribution with 5° bin width are calculated for each fluctuation angle, and Gaussian fitting $ e^{-\sigma*\rm{angle}^2} $ is performed to each distribution accordingly with $\sigma$ being the fitting parameter later used to quantify the peak width. The choice of the fitting equation is justified by the analytical prediction of energy dependence on deviation angles (to the leading order), which is detailed in the Results section.
In the case of narrow orientational distributions, the visualization is cropped to a smaller angle range after calculation performed in the full [-90°,90°] range.
In the case of planar rods and the longer homeotropic rods, it is impractical to do the re-slicing given the limited number of images taken, and the distributions found within achiral nematic LCs are adopted instead as we expect no difference between the horizontal and vertical distributions in such condition with no biaxiality.
The same histograms are subsequently utilized in the computation of the colloidal orientation order parameters, which is summarized in Results.

\subsection{Computer simulation of perturbed order of the molecular LC host around the colloidal particles}
\begin{figure*}
    \includegraphics[width=1.9\columnwidth]{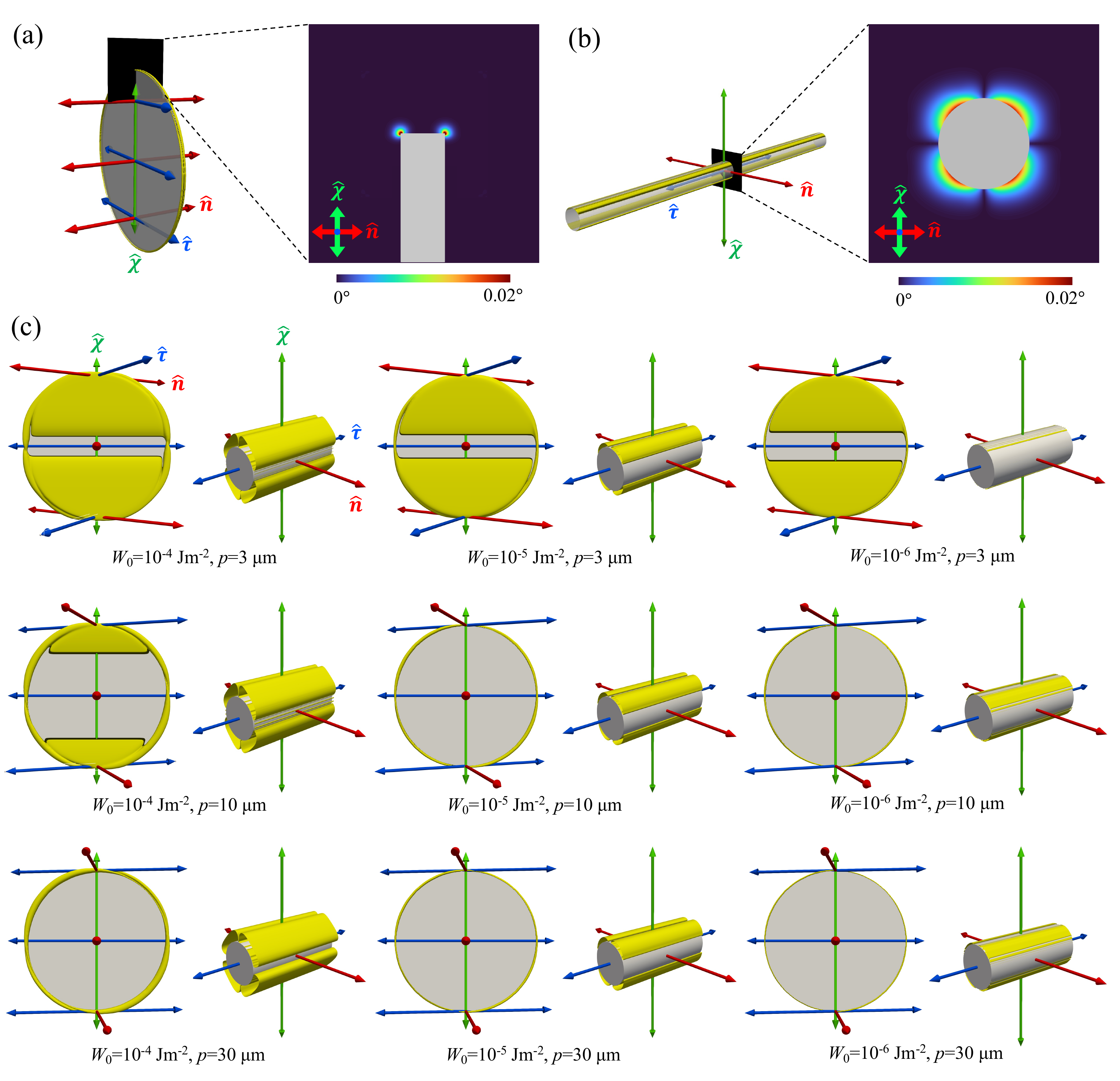}
    \caption{(a)-(b) Computer simulations of a thin colloidal disk (a) and a slender rod (b) immersed in a LC with weak chirality. Yellow contour surfaces mark the region where LC director deviations for $\bn$ (red axis) are 0.01° from its ideal helical state with no colloids present. The sectional areas perpendicular to $\btau$ (blue axis) are colored by the deviation angle as shown in the color scale. Homeotropic anchoring $W_0=10^{-6} \rm{J m^{-2}}$ and a helical pitch $p=30 \um$ are used for both simulations. (c) Simulations of colloidal disks and rods in energy-minimizing orientations within chiral LCs using various anchoring strengths $W_0$ and pitch lengths $p$, with values labeled for each simulation. Yellow contours enclose regions with director distortions larger than or equal to 0.1°, showing different levels of weak biaxiality. Rods in (c) are cropped for clarity. Axes defining the molecular frame are colored as in \fig{chiral_schematic}. Disk width $D_{\rm{c}}=1 \um $ and rod length $L_{\rm{c}}=1.7 \um$ for all simulations.}
    \label{numerical_visualization}
\end{figure*}
Computer simulations are carried out to study the interplay between molecular LC order near the colloidal surface and the colloidal orientation. The simulations are based on minimizing the mean-field Landau-de Gennes free energy for the molecular LC host \cite{mundoor2021,ravnik2009landau,mori1999multidimensional,tai2020surface,yuan2018chiral}.
We consider a thermotropic bulk free energy density describing the isotropic-nematic transition of LCs complemented with elastic contributions associated with LC director distortions occurring in the bulk volume of the LC\hide{ in the vicinity of the colloidal surface}:
\begin{align}
    f_{\rm {bulk}}^{\rm LC} &= \frac{A}{2} \bQ_{ij} \bQ_{ji} 
    + \frac{B}{3} \bQ_{ij} \bQ_{jk} \bQ_{ki} 
    + \frac{C}{4} ( \bQ_{ij} \bQ_{ji} )^2 
    \nonumber \\ &
    + \frac{L_1}{2} \left ( \pdQm{ij}{k} \right ) ^2 
    + \frac{L_2}{2} \pdQm{ij}{j} \pdQm{ik}{k} 
    \nonumber \\ &
    + \frac{L_3}{2} \pdQm{ij}{k} \pdQm{ik}{j} 
    + \frac{L_4}{2} \epsilon_{ijk} \bQ_{il} \pdQm{kl}{j}
    \nonumber \\ &
    + \frac{L_6}{2} \bQ_{ij} \pdQm{kl}{i} \pdQm{kl}{j}
    \label{LdGbulk}
\end{align}
with the 3-by-3 matrix $\bQ$ being the molecular tensorial order parameter describing the local average molecular ordering, $x_i$ ($i=1-3$) being cartesian coordinates, and $\epsilon$ the 3D Levi-Civita tensor. Summation over all indices is implied. Among the bulk energy terms, $A$, $B$, and $C$ are thermotropic constants and $L_i$ ($i=1-4,6$) are the elastic constants related to the Frank-Oseen elasticities via:
\begin{align}
    L_1 &= \frac{2}{27 (\Seq)^2} \left (K_{33} - K_{11} + 3 K_{22} \right ) \nonumber \\
    L_2 &= \frac{4}{9 (\Seq)^2} \left ( K_{11} - K_{24} \right ) \nonumber \\
    L_3 &= \frac{4}{9 (\Seq)^2} \left ( K_{24} - K_{22} \right ) \nonumber \\
    L_4 &= \frac{8}{9 (\Seq)^2} K_{22} \frac{2\pi}{p} \nonumber \\
    L_6 &= \frac{4}{27 (\Seq)^3} \left ( K_{33} - K_{11} \right ) 
    \label{K2L}
\end{align}
with $K_{11}$, $K_{22}$, $K_{33}$ and $K_{24}$ respectively denoting the splay, twist, bend and saddle-splay elastic moduli, and $\Seq$ being the equilibrium uniaxial scalar order parameter.
In addition to the bulk LC energy there is a contribution due to the boundary condition of the molecular LC at the colloidal surfaces which reads: 
\begin{align}
    f_{\rm {surf}}^{\rm LC} = W_{0} \left ( {\bf P}_{ik} \tilde{\bf Q}_{kl} {\bf P}_{lj} - \frac{3}{2} \Seq {\cos}^2 \theta_{\rm e} {\bf P}_{ij} \right )^2
    \label{LdGsurf}
\end{align}
with $W_{0}$ the surface anchoring strength, ${\bf P} = \bv \otimes \bv$ the surface projection tensor, $ \bv$ the surface normal director, and $\tilde{\bf Q} = \bQ + \frac{1}{2} \Seq {\bf I} $.
The equilibrium angle $\theta_{\rm e}=0$ corresponds to vertical or homeotropic anchoring at the boundary, and $\theta_{\rm e}=\pi$ leads to planar degenerate anchoring \cite{zhou2019degenerate}.
The dimensions and anchoring forces associated with colloidal particles are represented as boundary conditions inside the numerical volume with the parameters kept constant for each simulation.
Specifically, width $D_{\rm{c}}=1 \um$ and thickness $L_{\rm{c}}=10 \rm{nm}$ are used for thin disks in all computer simulated results, while $D_{\rm{c}}=28 \rm{nm}$ and $L_{\rm{c}}=1.7 \um$ for long rods, if not specified otherwise.
The total energy is then given by the integration of \eq{LdGbulk} over LC volume and \eq{LdGsurf} over colloid-molecule interfaces, with the colloidal volume excluded in the integral of free energy densities.

The total energy is numerically minimized based on the forward Euler method integrating:
\beq
    \frac{d\bQ}{dt}=-\frac{d F_{\rm total}^{\rm LC}}{d \bQ}
\eeq
with $t$ being the scaled energy-relaxation time of the LC. Adaptive Runge-Kutta method (ARK23) and FIRE, Fast Inertial Relaxation Engine, are adopted to increase numerical efficiency and stability \cite{kutta1901beitrag,sussman2019fast}.
The steady-state and termination of simulation are determined by the change in total free energy in each numerical iteration, which is usually monotonic decreasing. 
The values of biaxiality and local orientations of molecular directors are subsequently identified as eigenvalues and eigenvectors of $\bQ$ \cite{mottram2014introduction}. 
Alternatively, chirality axes ($\bn$,$\bchi$,$\btau$) are also represented as the eigenpairs of handedness tensor (or chirality tensor,  \cite{beller2014geometry,efrati2014orientation}) with results having high consistency with the biaxial approach mentioned above.
A more detailed description and comparison of the two approaches are to be given in Discussion.

For a thin homeotropic rod, the anchoring effect at the two ends of the particle is ignored in our simulations by setting the length of the simulation box equal to the rod length $L_{\rm c}$, which hardly changes the energies but greatly improves the numerical stability.
The following parameters are used for all computer simulations: $A=-1.72 \times 10^5 \rm{J m^{-3}}$, $B=-2.12 \times 10^6 \rm{ J m^{-3}}$, $C=1.73 \times 10^6 \rm{J m^{-3}}$, $K_{11}=6 \rm{pN}$, $K_{22}=3 \rm{pN}$, $K_{33} = 10 \rm{pN}$, $K_{24} = 3 \rm{pN}$ and $\Seq=0.533$.
The simulations are carried out in a Cartesian colloidal frame using equidistant grid sets and are consistent with those based on a radial-basis-function approach performed within the molecular frame \cite{fornberg2015solving}.

\section{Results}

\subsection{Symmetry-breaking at single particle level}
The symmetry-breaking of the nematic colloidal geometry, induced by the twisted alignment of chiral molecules, can be revealed at the single particle level by visualizing the LC distortion field around a single colloidal particle [\fig{numerical_visualization}]. For cylinder-shaped particles dispersed in isotropic solvent, such as thin disks or slender rods in ethanol, a continuous rotational symmetry can be observed locally with the symmetry axis being the disk normal or the long axis of the rod because the other two orthogonal axes are geometrically equivalent.
When the cylindrical colloids are dispersed into a chiral nematic, however, the uniaxial symmetry is broken in view of the boundary condition at particle-molecule interfaces and the far-field helical configuration of the LC molecules.
Although the realigning effect induced by the homeotropic boundary conditions at the colloidal surfaces is rather weak with $W_0=10^{-6} \rm{J/m^2}$ and deviation angle $\ll 1^{\circ}$ [\fig{numerical_visualization}a,b], for example, it is evident that the rotational symmetry of the surface-defect-dressed cylindrical colloids becomes discrete (2-fold rotation) once the colloids are immersed in a chiral LC [\fig{numerical_visualization}c].
Clearly, stronger surface anchoring forces and higher chirality (shorter pitch) lead to the significantly stronger molecular LC distortion as well as the ensuing emergent biaxiality as shown by the computer-simulated distortion in nematic director.
Also, the single-particle symmetry-breaking is observed even when the helical pitch $p$ is much larger than the particle dimension, with the particle sizes around 1-2 $\um$ [\fig{numerical_visualization}c]. This demonstrates that the shape biaxiality of the dressed colloidal particle imparted by the molecular chirality of the host is unavoidably developed at all strengths of surface anchoring and values of molecular chirality [\fig{numerical_visualization}c].
\hide{
even at weak surface anchoring conditions that are prevalent in our experiments.}

\subsection{Orientational distribution of the colloidal particles}
\begin{figure*}
	\includegraphics[width=1.9\columnwidth]{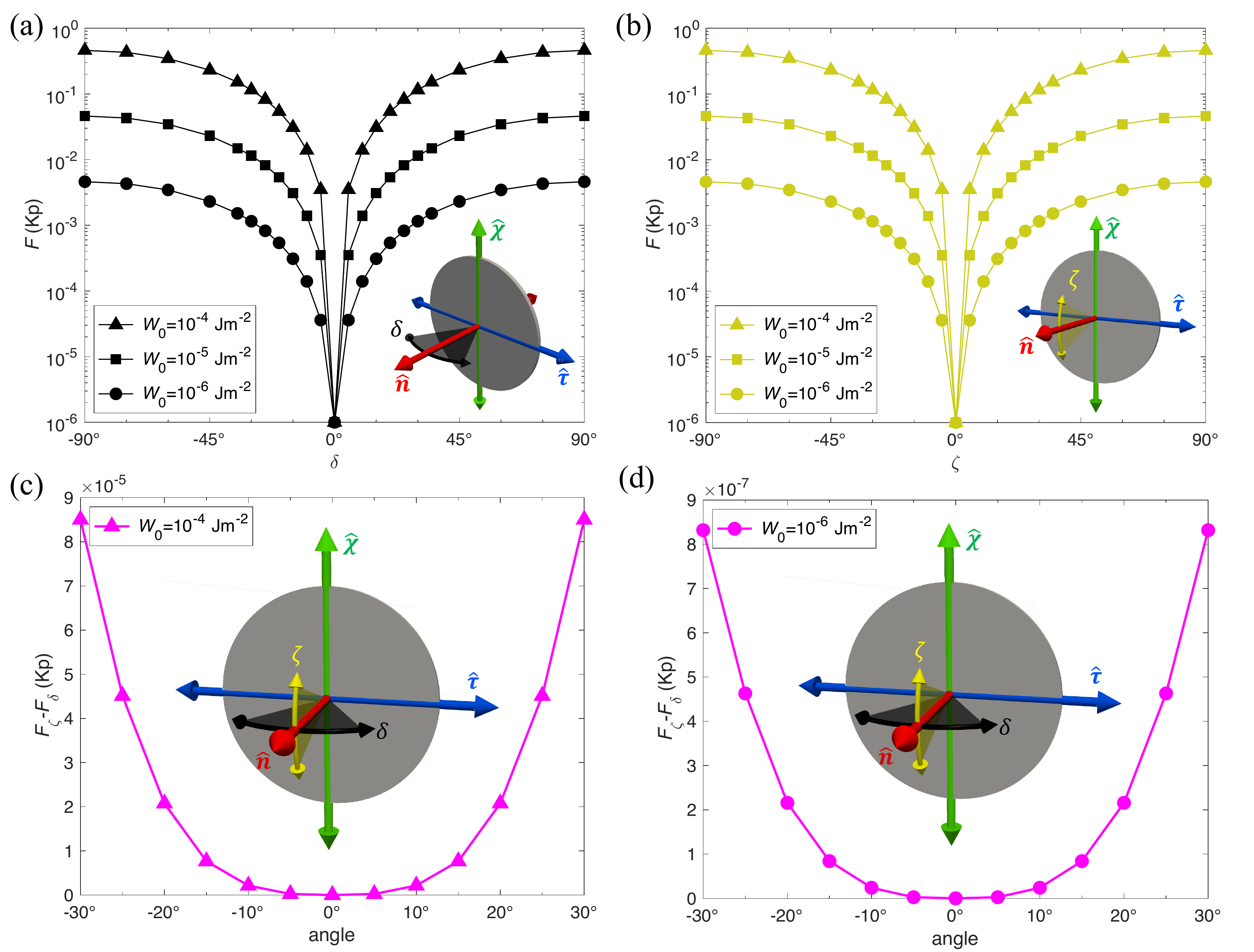}
	\caption{(a) Computer-simulated free energy of molecular chiral LC in the presence of a homeotropic disk at different surface anchoring strengths $W_0$ as a function of the azimuthal angle $\delta$ describing a rotation of the disk normal about the pitch axis $\bchi$ (green) defined in the molecular frame (inset). Data points for $W_0=10^{-4}, 10^{-5}$, and $10^{-6} \rm{J/m^2}$ are marked with triangle, square, and circle respectively. The energy is scaled by $Kp$ where $K=5.6 \rm{pN}$ is the average elastic constant and $p=30 \um$ is the pitch of cholesterics. (b) Numerical free energy profile for a homeotropic disk rotated about $\btau$ (blue axis). (c)-(d) The energy difference in (a) and (b) calculated for surface anchoring strengths $W_0=10^{-4} \rm{J/m^2}$ (c) and $10^{-6} \rm{J/m^2}$ (d), respectively. The lowest energies (disk normal aligned along red axis $\bn$) for each simulation set are chosen to be $10^{-6} Kp $ instead of 0 to avoid singularities when converting to a log-scale in (a) and (b). The axes in the insets define the molecular frame and are colored as in \fig{chiral_schematic}. Cholesteric pitch $p=30 \um$ for all simulations.}
	\label{numerical_disk}
\end{figure*}
To analyze the equilibrium orientation of the cylindrical particles, we perform several sets of simulations at various colloidal orientations and resolve the corresponding free energies.
A thin disk with perpendicular boundary condition [\fig{numerical_disk}], for example, favors alignment in which the disk normal vector orients along the molecular director $\bn$. Deviations away from the equilibrium direction give rise to an increase in the overall free energy of the system [\eq{LdGbulk} and \eq{LdGsurf}].
We emphasize that the free energy profiles are distinct for the two deviation angles $\delta$ and $\zeta$ in \fig{numerical_disk} (with lower energy penalty for orientational fluctuation along $\delta$). Though weak, the difference between the two angles and the broken uniaxial symmetry as a consequence of the chirality in the molecular LC host are unambiguous. 
Furthermore, \fig{numerical_disk} (c,d) demonstrate that a stronger surface anchoring force with a higher value of $W_0$ leads to more pronounced energetical nondegeneracy of the two deviation angles.
Using mean-field numerical simulation of the LC host, we are able to validate the local biaxial symmetry of the orientational probability distribution of an individual colloid, arising from the inequivalence of $\bchi$ and $\btau$ in the molecular LC host. \hide{More detailed characterizations of the orientational symmetry-breaking are given in the following sections.}

\begin{figure*}
	\includegraphics[width=1.9\columnwidth]{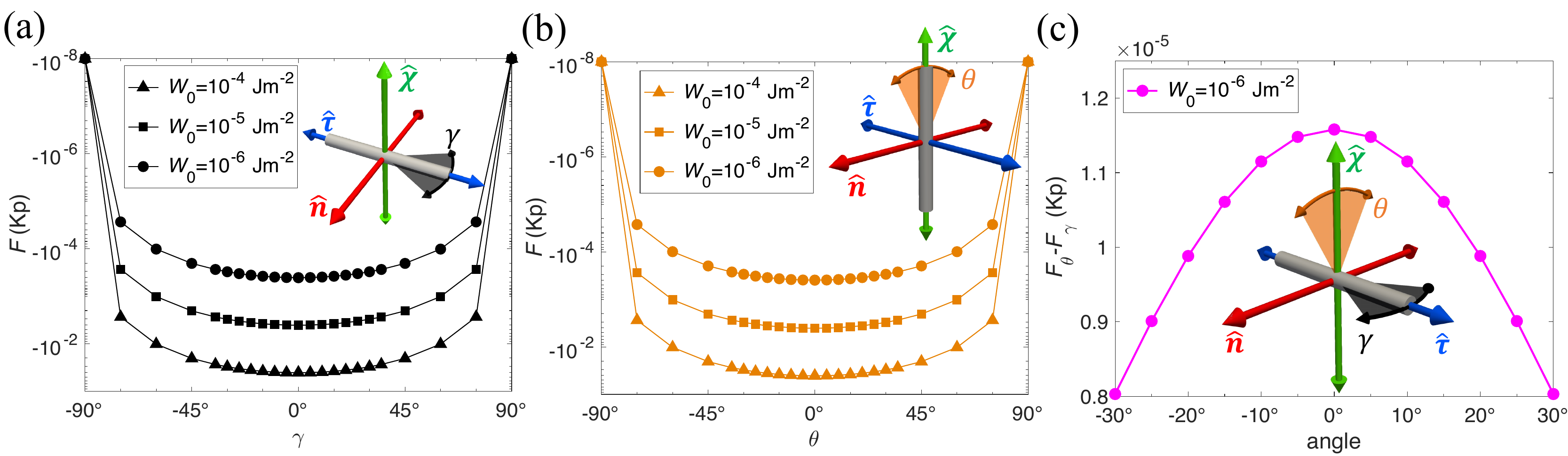}
	\caption{(a) Computer-simulated free energy of chiral 5CB-based LC surrounding a homeotropic rod at different surface anchoring strengths, $W_0$, and the azimuthal angle, $\gamma$, defined in the inset. (b) Simulated free energy of rods with different values of polar angle $\theta$. (c) The difference in energy profiles of homeotropic rods rotated along $\btau$ and $\bchi$ axis, as shown in the inset, simulated using anchoring strength $W_0=10^{-6}$. The case of rods aligning long $\bn$ (red axis) with the highest energy cost is taken as an energy reference point for each value of $W_0$, while the free energy value is chosen to be $-10^{-8} Kp$ instead of 0 to avoid singularities when converting to a log-scale in (a) and (b). The axes in the insets define the molecular frame and are colored as in \fig{chiral_schematic}. Cholesteric pitch $p=30 {\rm \um}$ for all simulations.}
	\label{numerical_rod}
\end{figure*}
In contrast to the case of a disk, a homeotropic rod feels a strong energy penalty when aligned towards $\bn$ and reaches a state of minimal surface anchoring energy when the long axis points along the $\btau$-direction such that the LC director at the rod surface naturally complies with the homeotropic surface anchoring condition [\fig{numerical_rod}] \cite{mundoor2018,tkalec2008interactions}. 
The symmetry-breaking of $\bchi$ and $\btau$, evident from the difference between the two energy landscapes [\fig{numerical_rod}], is observed again with deviation along the angle $\gamma$ being more energetically favored than that along $\theta$.
With the cases of colloidal rods along $\bn$, $\bchi$, and $\btau$ all giving distinct free energies, the biaxiality in the ensuing colloidal orientation probability distribution is explicit, which is contributed solely by the chirality in the 5CB host.
The results are in agreement with the empirical evidence shown below.

\subsection{Experimental observation and analysis of the colloidal orientation}
\begin{figure*}
	\includegraphics[width=1.9\columnwidth]{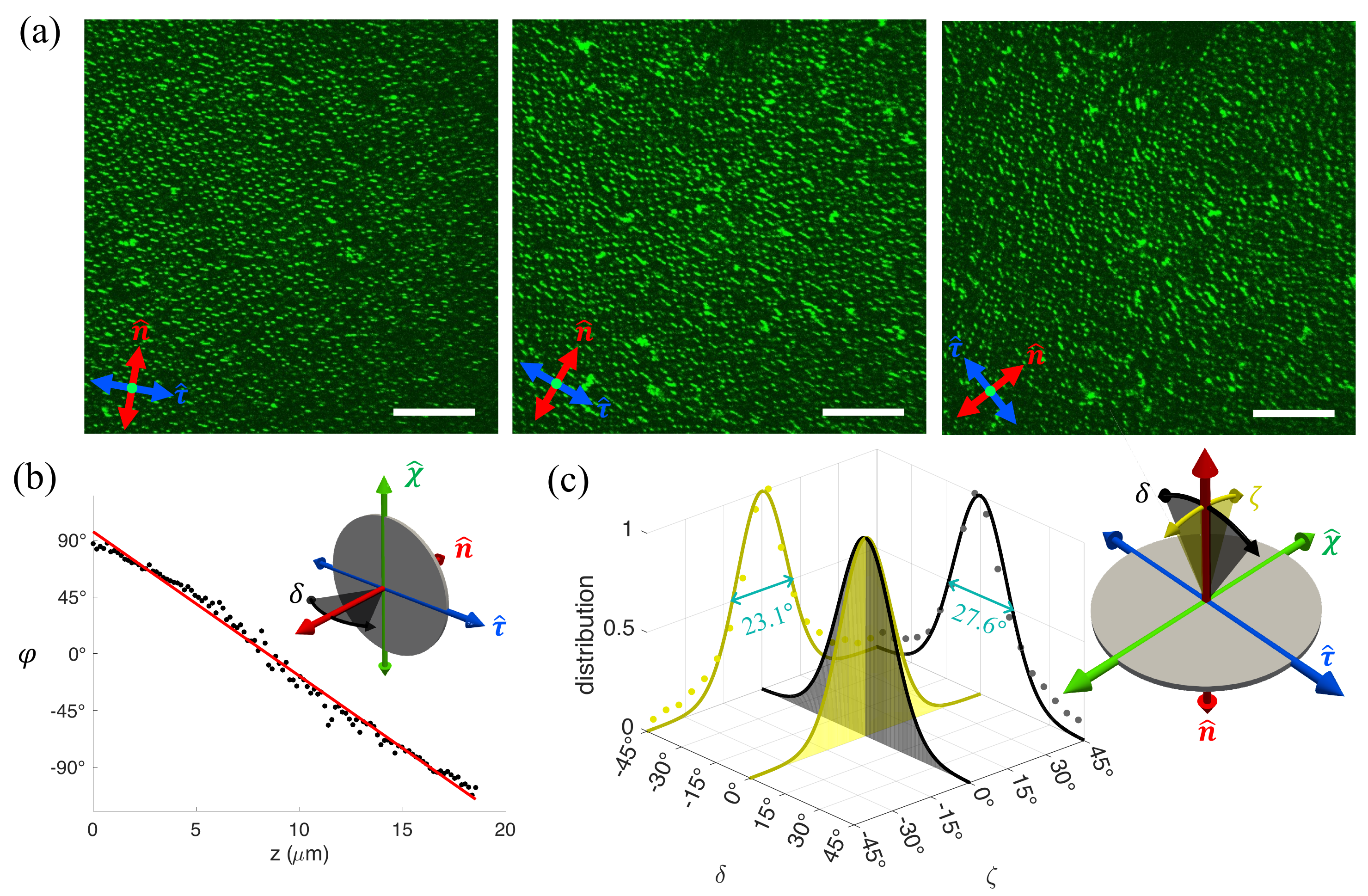}
	\caption{(a) Luminescence confocal images of homeotropic disks dispersed in chiral LC taken perpendicular to $\bchi$ with the molecular frame ($\bn$ and $\btau$) marked in each depth slice. (b) The azimuthal angle of the disk normal orientation $\varphi$ at different depth $z$ obtained from the same sample. Dots are the average values in each $z$ slice, and their linear fit is given by the red line, revealing the LC pitch $p \approx 30\um$. The inset illustrates the observed orientational fluctuation of disks (black double arrow) in the molecular frame ($\bn$, $\bchi$, $\btau$). (c) Disk azimuthal orientational fluctuation $\delta=\varphi-qz$ and polar orientational fluctuation $\zeta=\pi/2-\theta$ within the molecular frame (insets), with $\delta=0$ or $\zeta=0$ correspond to orientation along $\bn$ (red axis). Scale bars are $30 {\rm \um}$.}
	\label{exp_homeo_disk}
\end{figure*}
The alignment of the colloidal particles with respect to the helical director field of the molecular host LC is probed optically in our experiments.
For example, \fig{exp_homeo_disk}. (a) demonstrates confocal fluorescence micrographs of disk-shaped particles immersed in a 5CB liquid crystal doped with a chiral dopant. 
The molecular orthogonal frame ($\bn$,$\bchi$,$\btau$), which is marked in each micrograph, is robustly controlled by substrates with planar anchoring force (Methods).
The average normal direction of the colloidal disks in each vertical slice, which is expected to lie parallel to $\bn$, rotates along the sample depth, as clearly shown with the edge-on perspective [\fig{exp_homeo_disk}. (a)].
Subsequently, the twisted arrangement of the disk direction is analyzed and a quasi-uniform twisting rate is found throughout the sample depth [\fig{exp_homeo_disk}. (b)], with thermal fluctuation present.
We assume that the helix of molecular director $\bn$ has a linear trend identical to the one found using colloidal orientations, which is shown by the red line in \fig{exp_homeo_disk}. (b), since the period of the twisted arrangement of the colloids closely matches the designed molecular pitch $p$.
We also ascertain that the colloidal density remains very low such that the molecular LC alignment is not expected to be disturbed by the introduction of colloidal particles. 
Once the orientational distribution of the thin disks is projected onto the co-rotating molecular frame, we observed Gaussian-like distributions [\fig{exp_homeo_disk}. (c)]. 

With a particle number density (volume fraction $\approx$ 0.026\%) far below the phase transition threshold \cite{mundoor2021}, direct interactions between colloidal disks are negligible and each particle experiences an orientational potential imposed mainly by the surrounding molecular LC, as designed in our numerical simulations discussed above [\fig{numerical_disk}]. The statistical results of particles can thus be treated as the thermal distribution of a single particle and qualitative agreement is found when compared to the modeling [\fig{numerical_disk}].
As shown by the numerical modeling earlier, the colloidal director distributions are weakly asymmetric due to the biaxiality imparted by the chiral molecular host, demonstrating a stronger energy barrier of the thin disk fluctuating in the $\zeta$ direction.
As a result, the peak widths (full width at half maximum, FWHM) of the colloidal orientation distributions differ along two deviation angles (27.6° in $\delta$ and 23.1° in $\zeta$, \fig{exp_homeo_disk}. (c)), indicating a biaxial $D_2$ symmetry of an individual disk with 2-fold rotations around $\bn$ instead of uniaxial $D_{\infty}$ one, even though the cylindrical particles themselves are of uniaxial symmetry.

\begin{figure*}
	\includegraphics[width=1.9\columnwidth]{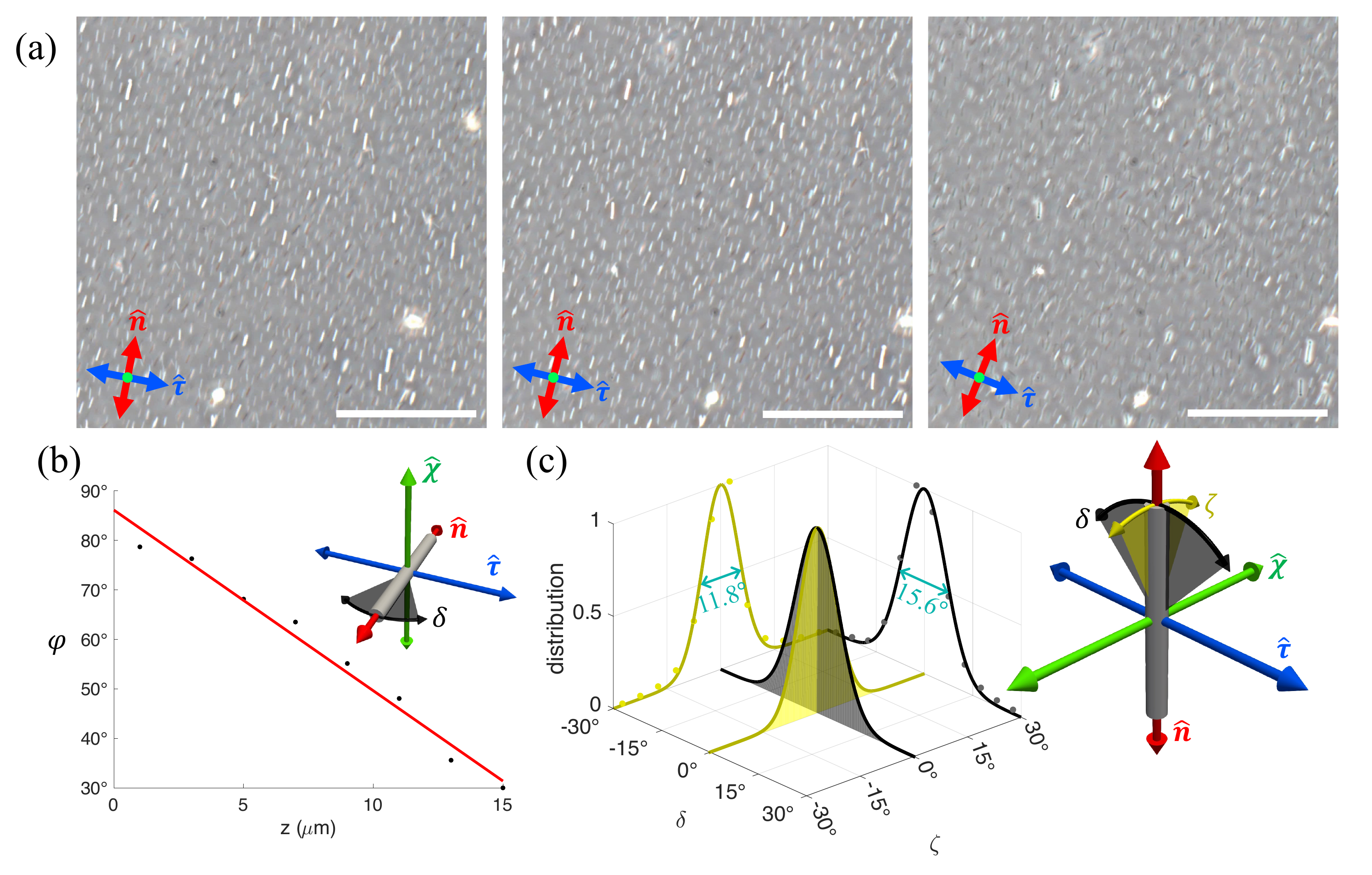}
	\caption{(a) Phase contrast micrographs showing depth slices of planar rods dispersed in chiral molecular 5CB. (b) Average rod orientation in each $z$-slice (dots) and linear fit (red line). The cholesteric pitch is $p \approx 100 \um$. (c) Azimuthal and polar orientational distribution of the rods, with $\bn$ (red axis) most populated in the molecular frame (insets). Scale bars are $30 {\rm \um}$.}
	\label{exp_planar_rod}
\end{figure*}
We further use phase contrast microscopy to determine the colloidal alignment directions for planar rods [\fig{exp_planar_rod}. (a)]. In this case, the colloidal thin rods with a planar surface condition are dispersed within the 5CB chiral molecular host. By carefully changing the focal position inside the sample, we obtain a good linear relationship between the average rod direction and depths, representing a helical structure within which the colloidal rods point along $\bn$ [\fig{exp_planar_rod}. (b)]. 
Like the case of homeotropic disks, we again clearly observe a biaxial orientational symmetry in the molecular frame revealed by distinct probability distributions along $\btau$ and $\bchi$ (FWHM=15.6° in $\delta$ and 11.8° in $\zeta$) despite the weak chirality ($p \approx 100 \um$) of the hybrid LC system [\fig{exp_planar_rod}. (c)].
The explicit experimental observations of the biaxial symmetry in the colloidal orientation probability distribution, in agreement with the numerical simulation, can only be attributed to the chiral twisting of the surrounding molecular LC.

\begin{figure*}
	\includegraphics[width=1.9\columnwidth]{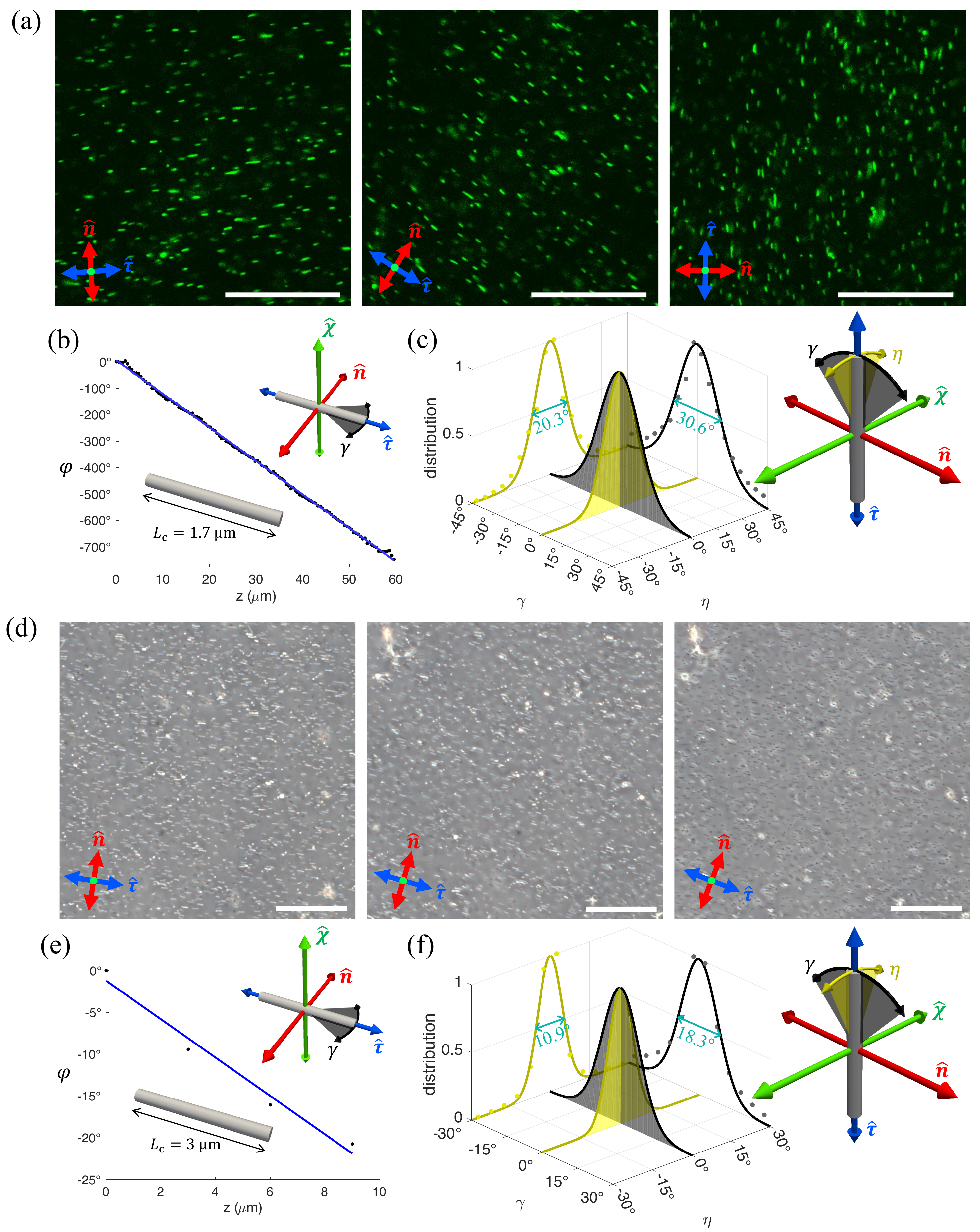}
	\caption{(a) Depth slices of homeotropic rods dispersion in a chiral LC taken using luminescence confocal microscopy. (b) The average orientation of the long axis of each rod $\varphi$ in each depth $z$ slice (dots) and its linear fit (blue line). LC pitch $p \approx 30 {\rm \um}$ and average rod length $L_{\rm c}=1.7 {\rm \um}$ (c) Orientational fluctuation of the rods measured in the LC molecular frame (inset), with the average direction being $\btau$ (blue axis). (d)-(f) Another sample of thin homeotropic rod dispersion in 5CB with cholesteric pitch $p \approx 150 \um$ and colloidal rod length $L_{\rm c}=3.0\um$ analyzed using phase contrast micrographs. All scale bars are $30 \um$.}
	\label{exp_homeo_rod}
\end{figure*}
To see the symmetry-breaking behavior in another type of alignment, such as perpendicular, we adopt the same methods of colloidal orientation analysis but for the colloidal inclusion of thin rods with a perpendicular boundary condition [\fig{exp_homeo_rod}].
In consistency with the numerical calculation [\fig{numerical_rod}], the rods aligns towards $\btau$ axis in thermal equilibrium [\fig{exp_homeo_rod} (a,d)].
After measuring the azimuthal angle $\varphi$ of rod long axis in each verticle frame and converting them to 3D distribution in the molecular orthogonal frame, we clearly see a narrower distribution is discovered for the longer rods [\fig{exp_homeo_rod} (c,f)]. 
Due to the larger surface area of the longer particle, to which the surface energy is proportional, a stronger energy barrier develops for the longer colloidal rods to fluctuate away from the energetically ideal configuration along $\btau$.
Furthermore, to our surprise, the orientational distributions of homeotropic rods behave dramatically differently from the non-chiral limit, in which case a degeneracy of alignment along the $\bchi$ and $\btau$-axes would be expected and the distribution along $\eta$ should be uniform to have the uniaxial symmetry.
Instead, we find an exceptionally strong energy barrier for rods deviating along $\eta$ towards the helical axis $\bchi$, with the Gaussian fittings showing peaks even sharper than those along $\gamma$ [\fig{exp_homeo_rod} (c,f)]. 
The symmetry of the hybrid LC system is thus strongly biaxial as experimentally illustrated by the distinct peak widths of the orientational probability distributions. 
We will demonstrate that the rod orientation distributions can not be interpreted from surface anchoring effects only, but can be explained by considering the elastic distortions generated in the bulk of the molecular host. This is addressed in detail in the following sections using a comprehensive analytical model. \hide{To further quantify the orientational energy potential of the colloids and validate the biaxiality of the orientation distribution, a more comprehensive analytical model is proposed below.}

\subsection{Analytical model}

\subsubsection{Surface anchoring free energy of a cylindrical disk immersed in a cholesteric host}

We consider a low-molecular-weight chiral liquid crystal with a director field $\bn(z)$ twisted along the $\bchi$-axis of a Cartesian laboratory frame that we denote by the normalized unit vectors $(\bx, \by, \bz)$ where $\bz$ coincides with the helical axis $\bchi$ in \fig{chiral_schematic}. The helical director field of a cholesteric, denoted by subscript ``$h$", may be parameterized as follows:
\beq
\bn_{h}(z) = \bx \cos q z  + \by \sin q z 
\label{ns}
\eeq
in terms of the cholesteric pitch $p = 2 \pi/q$ and handedness $q<0$ that we assume left-handed in agreement with experimental reality without loss of generality. Next, we immerse an infinitely thin cylindrical disk with aspect ratio $D_{\rm{c}}/L_{\rm{c}} \rightarrow \infty$ into a cholesteric host. The main symmetry axis of the colloidal disk is parameterized in the lab frame as $\bhu = \bx \sin \theta \sin \varphi + \by \sin \theta \cos \varphi + \bz \cos \theta $ in terms of a polar $\theta$ and azimuthal angle $\varphi$ with respect to the helical axis $\bz = \bchi$. The presence of the colloid will generate elastic distortions of the uniform director field $\bn_{h}(\bfr)$ due to the specific anchoring of the molecules at the colloidal surface, quantified by the surface anchoring strength $W_{0}>0$ (units energy per surface area). The extent of the elastic distortions around the colloid surface depends on the surface extrapolation length $\ell_{s} = K/W_{0}$ where $K$ denotes the average elastic constant of the thermotropic liquid crystal \cite{stark2001}. In our analysis, we first focus on the regime of infinitely large surface extrapolation length $ ( \ell_{s} \rightarrow \infty )$, in which case the elastic distortions around the immersed colloid are absent. For finite $\ell_{s}$, such as in the experimental situation, elastic distortions are weak but non-negligible and will be accounted for in the subsequent sections. If we assume the molecular director field $\bn$ to remain completely undistorted, the surface anchoring free energy can be obtained by using the Rapini-Papoular model for \eq{ns} and integrating over the colloid surface denoted by ${\mathcal S}$ \cite{rapini1969,senyuk2021nematoelasticity}:
\beq
F_{s} = -\frac{1}{2} W_{0} \oint d{\mathcal S}  (\bn_{h} \cdot \bv({\mathcal S}))^{2}
\label{rapo}
\eeq   
where $\bv$ represents a unit vector normal to the colloid surface in case of homeotropic (H) anchoring and tangential to the surface if the anchoring is planar (P).  
Let us denote its normal by $\bhu$ and ignore anchoring at the rim. We further define two unit vectors $\bhe_{1,2}$ orthogonal to the disk normal vector $\bhu$. The two principal anchoring scenarios, homeotropic (H) and planar (P), are expressed as follows:
\beq
\bv = \begin{cases}
      \bhu & \rm{H} \\
      \bhe_{1} \cos \xi + \bhe_{2} \sin \xi &  \rm{P} \\
       \end{cases}
      \label{plahom}
\eeq
The angle $0 < \xi < 2 \pi$ must be chosen randomly in the case when planar anchoring is degenerate across all directions on the disk surface, which is the case in the experimental situation. 

Ignoring finite-thickness effects for $L_{\rm{c}} \ll D_{\rm{c}}$ we then parameterize the face of the disk as follows:
\beq
\bfr_{{\mathcal S}} = \bfr_{0} + \frac{D_{\rm c}}{2} t [\bhe_{1} \sin \phi + \bhe_{2}\cos \phi ]
\eeq
with $0< t <1 $ and $0 < \phi < 2 \pi$. 
The surface anchoring energy per disk face is expressed as follows:
\begin{align}
F_{s} &= -\frac{1}{4}  W_{0} D_{\rm c}^{2} \int_{0}^{2 \pi} d \phi   \int_{0}^{1} dt t \int_{0}^{2 \pi} \frac{d \xi}{2 \pi} [ \bn_{h}( \bfr_{{\mathcal S}} \cdot \bchi ) \cdot \bv ]^{2}
\label{usurf2}
\end{align}
Leading to the following generic expression:
\begin{align}
 F_{s} = -\frac{\pi}{4}  W_{0}D_{\rm c}^{2} \left ( w_{1} + w_{2} \cos (2 \delta ) \frac{J_{1}(qD_{\rm c} | \sin \theta|)}{qD_{\rm c} | \sin \theta| } \right )
 \label{usp}
\end{align}
with $J_{1}(x)$ a Bessel function of the first kind, $\delta = \varphi  - q z$ the azimuthal angle with respect to the local cholesteric director, and coefficients:
\beq
w_{1} = \begin{cases}
    \frac{1}{2}  \sin ^{2} \theta    &  \rm{H} \\
      \frac{1}{8}( 3 + \cos ( 2 \theta ))   & \rm{P}
   \end{cases}
      \label{w1p}
\eeq
and
\beq
w_{2} = \begin{cases}
      \sin^{2} \theta    &  \rm{H} \\
    -\frac{1}{2}  \sin^{2} \theta    & \rm{P}
   \end{cases}
      \label{w2p}
\eeq
The surface anchoring strength of disks is expressed in dimensionless form by $\bar{W} = \beta W_{0}D_{\rm c}^{2}$ with $\beta^{-1} = k_{B}T$ the thermal energy in terms of temperature $T$ and Boltzmann's constant $k_{B}$. 
Taking disks with diameter $D_{\rm c} \approx 2 {\rm \um}$ and $W_{0} \approx 10^{-6} - 10^{-5} {\rm J m^{-2}}$ we find $\bar{W} \sim 10^{3}-10^{4}$, indicating that surface anchoring realignment is robust against thermal fluctuations in the experimental regime. 
For the case of homeotropic anchoring, the surface anchoring energy \eq{usp} reaches a minimum at an equilibrium angle $\theta^{\ast} = \pi/2 $ and $ \delta^{\ast} = 0$, demonstrating preferential alignment of the disk normal along the local LC host director $\bn$, in agreement with experimental observation [\fig{exp_homeo_disk}].

\subsubsection{ Surface anchoring free energy of a cylindrical rod immersed in a cholesteric host}

We may repeat the previous analysis to describe the case of a thin colloidal rod with $L_{\rm c} / D_{\rm c} \rightarrow \infty$ by neglecting small contributions associated with the ends of the cylinder so we only need to parameterize the cylindrical surface of magnitude $\pi L_{\rm c}D_{\rm c}$ following the principal contour $\bfr_{{\mathcal S}}(t) = \bfr_{0} +  \frac{L}{2}t\bhu $
with $-1<t<1$ of a cylinder with centre-of-mass $\bfr_{0}$. The surface anchoring free energy then becomes:
\beq
F_{s} = -\frac{1}{8} L_{\rm c} D_{\rm c} W_{0} \int_{0}^{2 \pi} d \phi  \int_{-1}^{1} dt  [ \bn_{h}( \bfr_{{\mathcal S}} \cdot  \bchi ) \cdot \bv ]^{2}
\label{usurf}
\eeq
In order to describe various anchoring situations we define two unit vectors $\bhe_{1,2}$ orthogonal to $\bhu$ and parameterize:
\beq
\bv = \begin{cases}
      \bhe_{1} \cos \phi + \bhe_{2} \sin \phi &  \rm{H} \\       
      -\bhe_{1} \sin \phi \cos \xi + \bhe_{2} \cos \phi \cos \xi + \bhu \sin \xi &  \rm{DP} \\
        \bhu & \rm{SP}
   \end{cases}
\eeq
In the case of homeotropic (H) anchoring the molecular director favors perpendicular alignment to the cylindrical surface, whereas for simple planar (SP) surface anchoring along the main rod direction is favored. For completeness, we also include the more general degenerate planar (DP) case where all anchoring directions perpendicular to the local surface normal are equally probable. In order to account for all possible rod orientations with respect to the molecular field, the angle $\xi$ can take values between 0 and $\pi$.
We obtain the following generic expression:
\begin{align}
 F_{s} = -\frac{\pi}{8} L_{\rm c}D_{\rm c}W_{0} \left ( w_{1} + w_{2} \cos (2 \delta )  \frac{\sin (qL_{\rm c} \cos \theta)}{qL_{\rm c}} \right )
 \label{us}
\end{align}
with $\delta = \varphi - qz $ the azimuthal angle along a particle frame co-rotating with the  helical director so that $ \int d \bhu = \int_{0}^{2 \pi} d \delta \int_{-1}^{1}  d ( \cos \theta )$  and $w_{1}$ and $w_{2}$ are angle-dependent coefficients that depend on the particular anchoring situation: 
\beq
w_{1} = \begin{cases}
      (1+ \cos^{2} \theta)  &  \rm{H} \\
      \tfrac{1}{2}(3 - \cos^{2} \theta ) &  \rm{DP}   \\
   2 \sin^{2} \theta & \rm{SP}
   \end{cases}
      \label{w1}
\eeq
and
\beq
w_{2} = \begin{cases}
    -  \sin \theta \tan \theta    &  \rm{H} \\
    \tfrac{1}{2} \sin \theta \tan \theta  &  \rm{DP} \\
    2 \sin \theta \tan \theta  & \rm{SP}
   \end{cases}
      \label{w2}
\eeq
in terms of the polar $\theta$ and azimuthal rod angle $\varphi$ with respect to the helical axis along the $\bchi$-direction.

  
For the homeotropic (H) case the free energy is minimal at an equilibrium angle $\theta^{\ast} = 0$ (with the azimuthal angle $\varphi$ randomly distributed) which corresponds to the rod being aligned along the $\bchi$ direction. However, there is a second, degenerate minimum at $\theta^{\ast} = \pi/2$ and $\delta^{\ast} = \pi/2$, that describes a rod pointing along the $\btau$-axis. The minimum surface anchoring energy is $F_{s} = -(\pi/4) L_{\rm c}D_{\rm c} W_{0}$ for both cases. The energy barrier between the two minima is only about 1 $k_{B}T$ per rod so thermal fluctuations should easily make the colloids switch from one state to the other while staying perpendicular to $\bn$. 
 For both simple planar (SP) and degenerate planar (DP) anchoring we only find a single minimum at $\theta^{\ast} = \pi/2$ and $\delta^{\ast} = 0 $, i.e., the rod preferentially aligns along the revolving local nematic director $\bn$ as observed in experiments [see \fig{exp_planar_rod}].

\subsubsection{Equilibrium colloid orientation}

\begin{figure*}
	\includegraphics[width=2\columnwidth]{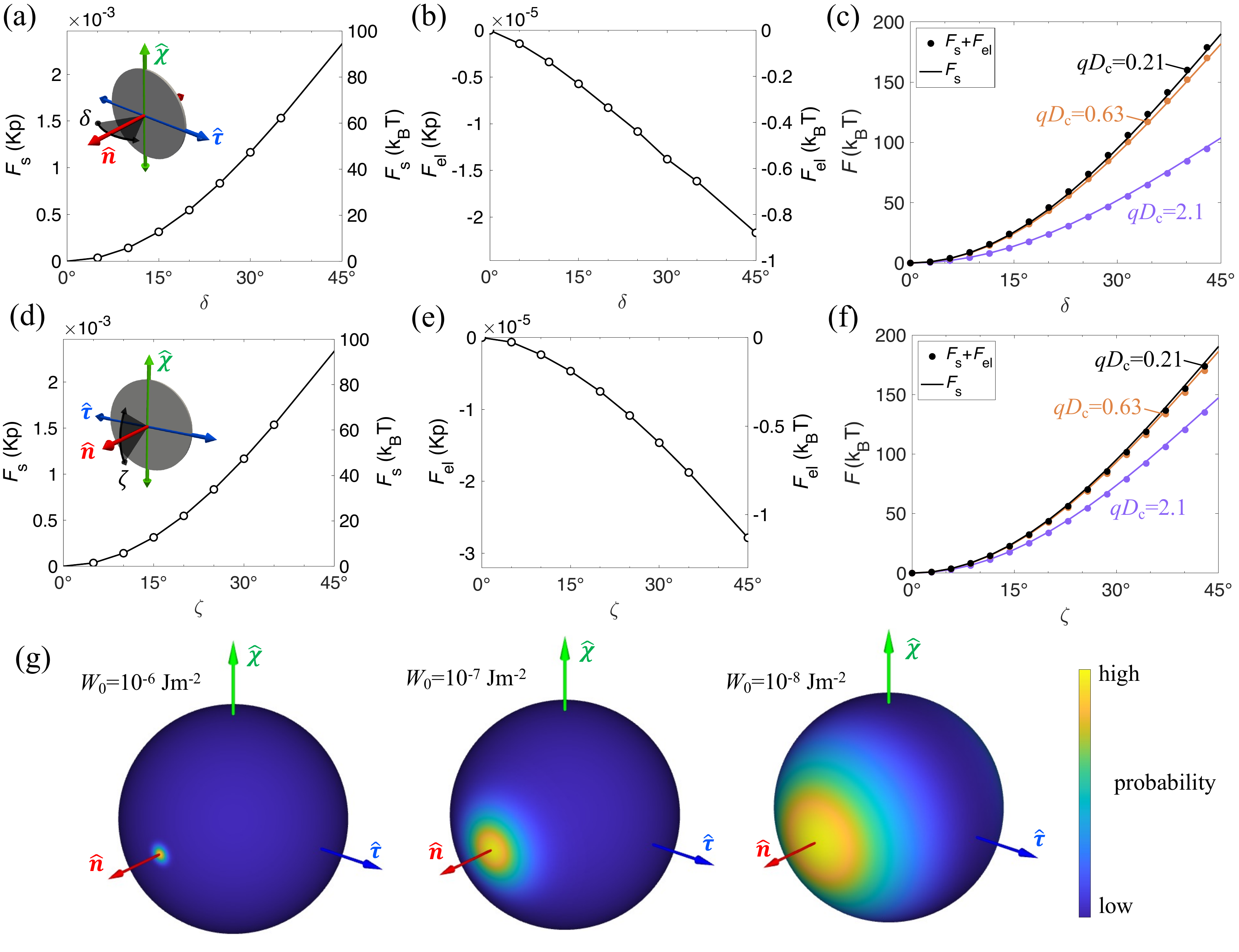}
	\caption{(a)-(b) Computer-simulated LC surface anchoring energy (a) and elastic distortion energy (b) using a Q-tensor description  of a chiral 5CB-based LC surrounding a homeotropic disk at different angles $\delta$ defined in the inset. The left and right axes provide different units of energy. (c) The prediction from analytical theory for different values of LC chiral strengths $qD_{\rm c}$. Solid lines correspond to the surface anchoring energy alone, while dots include the contribution of weak elastic distortion around the colloidal disk (see Appendix B). (d)-(f) Numerical simulation of surface energy (d) and elastic energy (e) and analytical prediction [\eq{usp}, Appendix B] (f) of the free energies for homeotropic disks at different angles $\zeta$ (defined in the inset). (g) Unit-sphere projections of the predicted local orientational probability of a disk immersed distribution in a chiral LC with various anchoring strengths $W_{0}$. Surface anchoring strength $W_{0}=10^{-6} \rm{J m^{-2}}$ for (a)-(f) and cholesteric pitch $p=30 {\rm \um}$, $qD_{\rm{c}}=0.21$ if not otherwise specified. Disk dimensions $L_{\rm{c}}=10 \rm{nm}$ and $D_{\rm{c}}=1 \um $ for all simulations and calculations. The energy zero points are chosen at $\delta=0$ or $\zeta=0$ for clarity.}
	\label{numerical_disk_detail}
\end{figure*}

Balancing the surface anchoring free energy with the orientational entropy of the individual colloids we easily establish the orientational probability distribution through the Boltzmann distribution:
  \beq
  f( \bhu ) = \mathcal{N} \exp ( - \beta F_{s} )
  \label{fsingle}
  \eeq
 with $\mathcal{N}$ a normalization constant ensuring that $\int d \bhu f(\bhu) = 1$. 
 It is easy to infer from \eq{usp} and \eq{us} that the polar and azimuthal angles are strongly coupled in general. This indicates that the local distribution of colloidal orientations around the principal alignment directions ($\btau$, $\bchi$ and $\bn$) [\fig{chiral_schematic}] is rendered {\em biaxial} by the chiral twist, in line with our experimental observations. Some examples of $f$ for disks are depicted in \fig{numerical_disk_detail} clearly demonstrating preferred alignments of the disk normals along $\bn$ (red arrow). 
We point out that the consistency found between the numerical modeling and the analytic theory is remarkable considering the complexity involved in computer simulation due to a wide range of length scales including molecular, colloidal (with high aspect ratios) and surface extrapolation lengths as well as the simplicity and approximations adopted in the analytic model.

The most interesting situation arises in the case of colloidal rods with homeotropic (H) 
 anchoring where there is a subpopulation of rods aligned along the helical axis ($qL_{\rm c}=1$). In order to gain further insight into the orientational symmetry of those rods, we perform a small-angle expansion around the equilibrium angle $\theta^{\ast} = 0$ and retain the leading order coupling term between the two principal angles $\theta$ and $\delta$. The angular fluctuations about the helical axis (green) are then described by the following free energy 
\beq
F_{s} \approx \frac{\pi}{8} L_{\rm c}D_{\rm c}W_{0}  j_{0}(qL_{\rm c}) \cos (2 \delta) \theta^{2}
\label{ht_fluc_new}
\eeq
with $j_{0}(x) = \sin (x)/x$. It suggests that the subpopulation of rods aligned along the helical axis in fact adopt a {\em twist-bend}-type organization with a pitch $q$ identical to that of the molecular host. Contrary to cholesterics, these phases are characterized by a nematic director co-aligning with the helical axis. However, the situation here is more subtle given that chirality is only manifested at the level of orientational fluctuations around a mean director ``backbone" that itself is not chiral.  We identify a further interesting feature; depending on the sign of $j_{0}(qL_{\rm c})$ the twist-bend helix may be either in phase with the molecular helix ($\delta_{e} =0$ for $qL_{\rm c}=4$) or out-of-phase ($\delta_{e} = \pi/2$ for $qL_{\rm c}=2$). In the next Section, we will show that the pure Rapini-Papoular description is inadequate in accounting for the experimental observations in \fig{exp_homeo_rod} for one of the experimental geometries (rods with perpendicular boundary conditions) and that weak elastic distortions around the colloidal rods must be accounted for to explain their strong preference for pointing along $\btau$.

\subsubsection{Elastic deformations surrounding the rod surface}

\hide{
The experiments involve rods with length $L_{\rm c}\approx 1.7 \um$, thickness $D \approx 25 {\rm nm}$ and  homeotropic (H) surface anchoring with a weak amplitude in the range $W_{0} \approx 10^{-4} - 10^{-7} {\rm J m^{-2}}$. From this we infer that ${\bar W} \gg 1$ indicating that the realigning torques generated by surface anchoring easily withstand thermal fluctuations. The pitch of the cholesteric host is about $p \approx 30 {\rm \um} $ so that $qL_{\rm c}  \approx 0.335 $. 
}

So far we have completely ignored the role of weak elastic deformations of the host director ($\ell_{s} = K/W_{0} \rightarrow \infty$) and assumed that the rod orientation is dominated entirely by surface anchoring effects. The experimental reality, however, is that the surface anchoring extrapolation length is large but finite ($\ell_{s} \approx 600 nm \gg D_{\rm c}$). Experimental observations compiled in \fig{exp_homeo_rod} point at a scenario where rods orient preferentially along the $\tau$ direction, rather than the helical axis ($\bchi$) as predicted from minimizing the bare Rapini-Papoular surface anchoring energy. A plausible reason as to why rod alignment along the helical axis ($\bchi)$ seems unfavorable is that it involves a twisting of the surface disclination that runs along the rod contour which costs elastic energy. No such twisting is required if the rod points along $\btau$. Clearly, the discrepancy between experiment and theory must be attributed to the elastic distortions running along the rod surface (and their subsequent twisting) which has been ignored in our considerations thus far. In principle, weak director distortions may also lead to a mild decrease in the bulk nematic order parameter, particularly in regions where the director curvature is strong. In our analysis, we will assume that the bulk scalar order parameter ($\Seq$) of the host is constant throughout the system. Even in the near-field limit close to the rod surface where deviations from bulk nematic order are strongest, we expect local distortions in bulk nematic order to be minor compared to the (infinitely) strong anchoring scenario that is considered in the theoretical study by Brochard and De Gennes \cite{brochard1970theory}.

\subsubsection{Elastic energy of a twisted disclination along the main rod direction}

We will now attempt to quantify the twisted disclination effect by introducing an angular deviation $\Phi(\bfr)$ and express the helical host director as follows:
\beq
\bn_{h}(\bfr) = \bx \cos( q z + \Phi(\bfr_{\perp}))  + \by \sin (q z + \Phi(\bfr_{\perp})) 
\label{nseblue}
\eeq
with $\bfr$ denoting a 3D distance vector and $\bfr_{\perp}$  the lateral distance perpendicular to the helical axis $\bchi$. The total free energy of a colloidal rod inclusion aligned along the helical axis is given by the Rapini-Papoular surface anchoring term \eq{rapo} combined with the Frank elastic free energy in the presence of chirality \cite{gennes-prost}:
\begin{align} 
F &= \tfrac{1}{2} \int d \bfr \left [ K_{11} (\nabla \cdot \bn_{h})^{2}  + K_{22} (\bn_{h} \cdot \nabla \times \bn_{h} + q)^{2} \right . \nonumber \\ 
& \left . +   K_{33} (\bn_{h} \times \nabla \times \bn_{h})^{2} \right ]  
-\tfrac{1}{2} W_{0} \oint d{\mathcal S}  (\bn_{h} \cdot \bv({\mathcal S}))^{2}
\end{align}
with $K_{11}$, $K_{22}$ and $K_{33}$ respectively denoting the splay, twist and bend elastic modulus, as defined in our simulation model in Section II. E. For simplicity, we ignore any contributions due to surface elasticity and assume the rod to be infinitely long and elastic distortions to occur only along the radial direction $\bfr_{\perp}$.
Employing cylindrical coordinates $\Phi(\bfr_{\perp}) = \Phi (r, \vartheta ) $, expanding up to second order in $q$ and integration over $\vartheta$ we obtain for the free energy $ F_{el}$ per unit rod length:
\begin{align}
 \frac{F_{el}}{L_{\rm c}} &=  \tfrac{1}{2} \int d  \bfr_{\perp} \left \{ \frac{K_{11}}{r^{2}} (1+ \partial_{\vartheta} \Phi)^{2} + K_{33} (\partial_{r} \Phi)^{2} \right . \nonumber \\ 
  & \left . + \frac{(qL_{\rm c})^{2}}{12} \Delta K  \left  [ \frac{1}{r^{2}} (1+ \partial_{\vartheta} \Phi)^{2} - (\partial_{r} \Phi)^{2} \right ] \right \} 
  \label{felo}
\end{align}
where $\Delta K = K_{33} - K_{11} >0 $ denotes the difference between the bend and splay moduli. The elastic anisotropy turns out to be of crucial importance since the twist correction $\mathcal{O}(q^{2})$ vanishes in case of the one-constant approximation $K_{11} = K_{33} = K_{22} = K$.
Similarly, the surface anchoring free energy reads up to quadratic order in $qL_{\rm c} \ll 1$: 
\beq
  \frac{F_{s}}{L_{\rm c}} =  -\frac{W_{0}}{2} \oint_{\mathcal{C}} d %
\vartheta \left \{ \cos^{2}(\vartheta - \Phi)  - \frac{(qL_{\rm c})^{2}}{12} \cos [2 ( \vartheta - \Phi)]  \right \}   \eeq
where $\mathcal{C}$ denotes the circular contour of the rod cross-section with diameter $D_{\rm{c}}$.
For weak distortions $\Phi \ll 1$ we linearize for $\Phi$ and obtain:
\beq
  \frac{F_{s}}{L_{\rm c}} \approx \frac{F_{s}^{(0)}}{L_{\rm c}} -\frac{W_{0}}{2} (1- \tfrac{1}{6} (qL_{\rm c})^{2}) \oint_{\mathcal{C}} d  \vartheta   \sin 2 \vartheta  \Phi 
  \label{fsdist}
\eeq
The first term is the contribution for the {\em undistorted} director field previously analyzed:
\begin{align}
  F_{s}^{(0)} &=  -\frac{L_{\rm c}W_{0}}{2} \oint_{\mathcal{C}} d %
\vartheta \left \{ \cos^{2}\vartheta - \frac{(qL_{\rm c})^{2}}{12} \cos 2  \vartheta \right \} \nonumber \\  
& \sim -\frac{\pi}{4} W_{0} L_{\rm c}D_{\rm c}
\label{sabasis}
\end{align}
which corresponds to \eq{usurf} for a homeotropic rod aligned perpendicular to the helical axis ($\theta = \delta = \pi/2$) in the large pitch limit $qL_{\rm c} \ll 1$. The second term in \eq{fsdist} accounts for the change of surface anchoring free energy generated by the elastic distortions.
The change of elastic free energy induced by the twist follows from:
\begin{align}
 \Delta F_{\rm twist}^{(el)}  &\approx  \frac{1}{24} (qL_{\rm c})^{2} L_{\rm c} \Delta K   \mathcal{F} [ \Phi_{0} ] 
 \label{Ftwistel}
\end{align} 
where $\Phi_{0}$ denotes the distortion angle for the {\em untwisted} system, and: 
\beq
 \mathcal{F} [ \Phi_{0} ]  = \int d \bfr_{\perp}  \left [ \frac{1}{r^{2}} (1+ \partial_{\vartheta} \Phi_{0})^{2} - (\partial_{r} \Phi_{0})^{2} \right ] 
 \label{fmans}
\eeq
is a dimensionless quantity measuring the extent of the surface disclination surrounding the cylinder. Applying the one-constant approximation which does not lead to qualitative changes in this context, we determine $\Phi_{0}$ from minimizing:    
\begin{align}
 \frac{F_{el}(q=0)}{KL_{\rm c}} &=  \tfrac{1}{2} \int d \bfr_{\perp}  \left \{ \frac{1}{r^{2}} (1+ \partial_{\vartheta} \Phi)^{2} + (\partial_{r} \Phi)^{2} \right \} 
\end{align}
so that $(\delta F_{el} / \delta \Phi )_{\Phi_{0}} = 0$ and $\ell_{s}  = K/W_{0}$ defines the (finite) surface anchoring extrapolation length.  Functional minimization of the free energy we obtain the Laplace equation in polar coordinates:
\beq
\partial_{r}^{2} \Phi_{0} + \frac{1}{r} \partial_{r} \Phi_{0}  + \frac{1}{r^{2}} \partial_{\vartheta}^{2}\Phi_{0} =0 
\label{lapo}
\eeq
subject to the boundary conditions: 
\begin{align} 
\Phi_{0}( \infty, \vartheta ) &= 0 \nonumber \\
\partial_{r} \Phi_{0}(D_{\rm c}/2, \vartheta) &= (4 \ell_{s})^{-1} \sin 2 \vartheta
\label{phioo}
\end{align}
with the latter denoting a Neumann boundary condition at the colloid surface imparted by surface anchoring contribution \eq{fsdist}. This ensures that the interior of the rod cross-section is excluded from the spatial integrations.
The result is a simple dipolar field:
\beq
\Phi_{0}(r , \vartheta ) = -\frac{D_{\rm c}}{16 \ell_{s}} \left ( \frac{D_{\rm c}}{2r}   \right )^{2} \sin 2 \vartheta 
\label{dipolarf}
\eeq
Plugging this back into \eq{fmans} and integrating we find that the difference in elastic energy between the twisted ($\bchi$) and untwisted ($\btau$) alignment directions in independent of the surface anchoring extrapolation length $\ell_{s}$ and increases logarithmically with system size $\ell_{\rm max}$:
\beq
 \Delta F_{\rm twist}^{(el)} \sim  \frac{\pi }{12} (qL_{\rm c})^{2} L_{\rm c} \Delta K  \ln \left ( \frac{2 \ell_{\rm max}}{ D_{\rm c}} \right )
 \label{twidi}
 \eeq
 Taking $\ell_{\rm max} = L_{\rm c}$ as typical size cut-off, a splay-bend elastic anisotropy $\Delta K =  4 {\rm pN}$ we find that $\Delta F_{\rm twist}  \sim \mathcal{O}(10^{2}  k_{B}T)$. 
The change in Rapini-Papoular surface anchoring free energy associated with a twist of the director distortions reads:
\begin{align}
  \Delta F_{\rm twist}^{(s)} & 
  \sim - \frac{\pi W_{0}L_{\rm c} D_{\rm c}}{92} \frac{D_{\rm c}}{ \ell_{s}}  (qL_{\rm c})^{2}
\end{align}
which is only a fraction of the thermal energy so that the total distortion-induced free energy change is estimated from $\Delta F_{\rm twist} \approx \Delta F_{\rm twist}^{(el)}$.

In Appendix A we discuss an analytical model that allows us to quantify the weak elastic distortions that occur when the rod remains perpendicular to the helical axis $\bchi$ but is allowed to display angular fluctuations in the $\bn-\btau$-plane, as illustrated by the angular probability distribution $f(\gamma)$ in \fig{numerical_rod_detail}. For disks, a similar model is discussed in Appendix B. There we demonstrate that the elastic distortions around the disk surface are intrinsically chiral but are very weakly developed and do not lead to qualitative changes in their realignment behavior as compared to predictions based on the Rapini-Papoular energy alone \eq{usp}.

\hide{

In order to explore the role of the cylindrical surface of the immersed rod we further consider the saddle-splay contribution which reads:
\beq
F_{se} = -\frac{K_{24}}{2} \int d \bfr   \nabla \cdot ( \bn_{h} \nabla \cdot \bn_{h} +  \bn_{h} \times \nabla \times \bn_{h} )
\eeq
Similar to the bulk elastic expression, we may expand for weak twist and use the dipolar solution \eq{dipolarf} to express the free energy change associated with saddle-splay elasticity upon twisting the surface defect. After elaborate rearrangements we find:
\beq
\Delta F_{\rm twist}^{se} \approx \frac{7 \pi}{12288} K_{24} L (qL_{\rm c})^{2} \left ( \frac{D}{\ell_{s}} \right )^{2}
\eeq
}

\subsubsection{Effective realigning potential per rod}

\begin{figure*}
	\includegraphics[width=2\columnwidth]{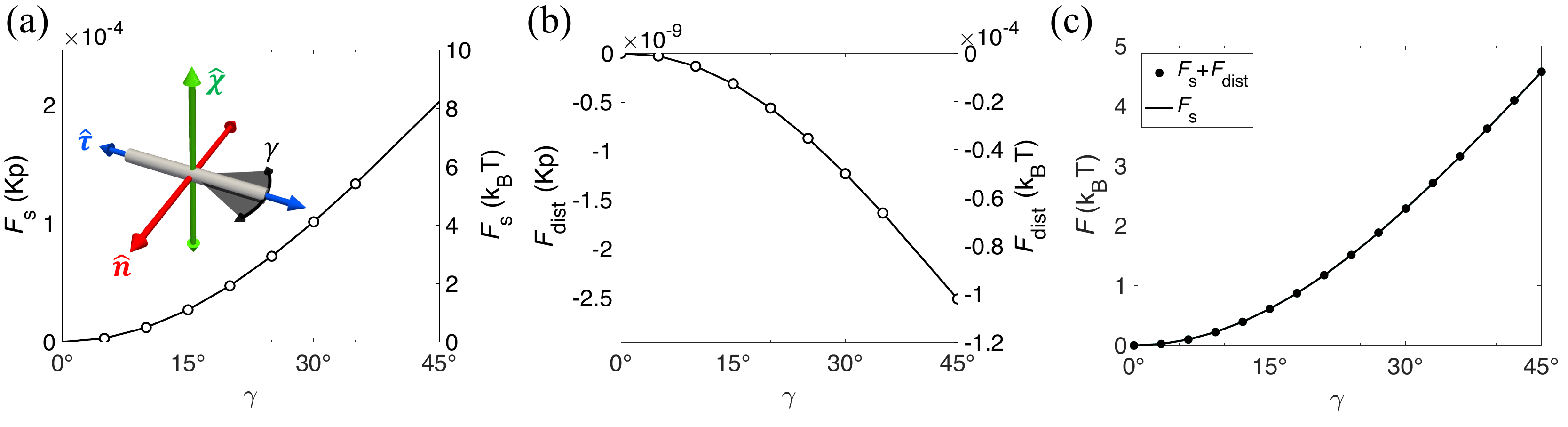}
	\caption{(a)-(b) Numerical LC surface anchoring energy (a) and elastic distortion energy (b) with homeotropic rods at different angles $\gamma$ defined in the inset. (c) The corresponding theoretical values with [\eq{frodinterp}] or without [\eq{us}] elastic distortion. \hide{(d)-(f) Numerical simulation of surface energy (d) and elastic energy (e) and theory prediction (f) of free energies for homeotropic disks at different angles $\theta_n$ (defined in the inset).} Surface anchoring strength $W_{0}=10^{-6} \rm{J m^{-2}}$, pitch $p=30 \um$, rod size $\Lc=1.7\um$ and $\Dc=28 \rm{nm}$ are used for all simulations and calculations. The energy zero points are chosen at $\gamma=0$ for clarity.}
	\label{numerical_rod_detail}
\end{figure*}

\begin{figure*}
    \includegraphics[width = 2\columnwidth]{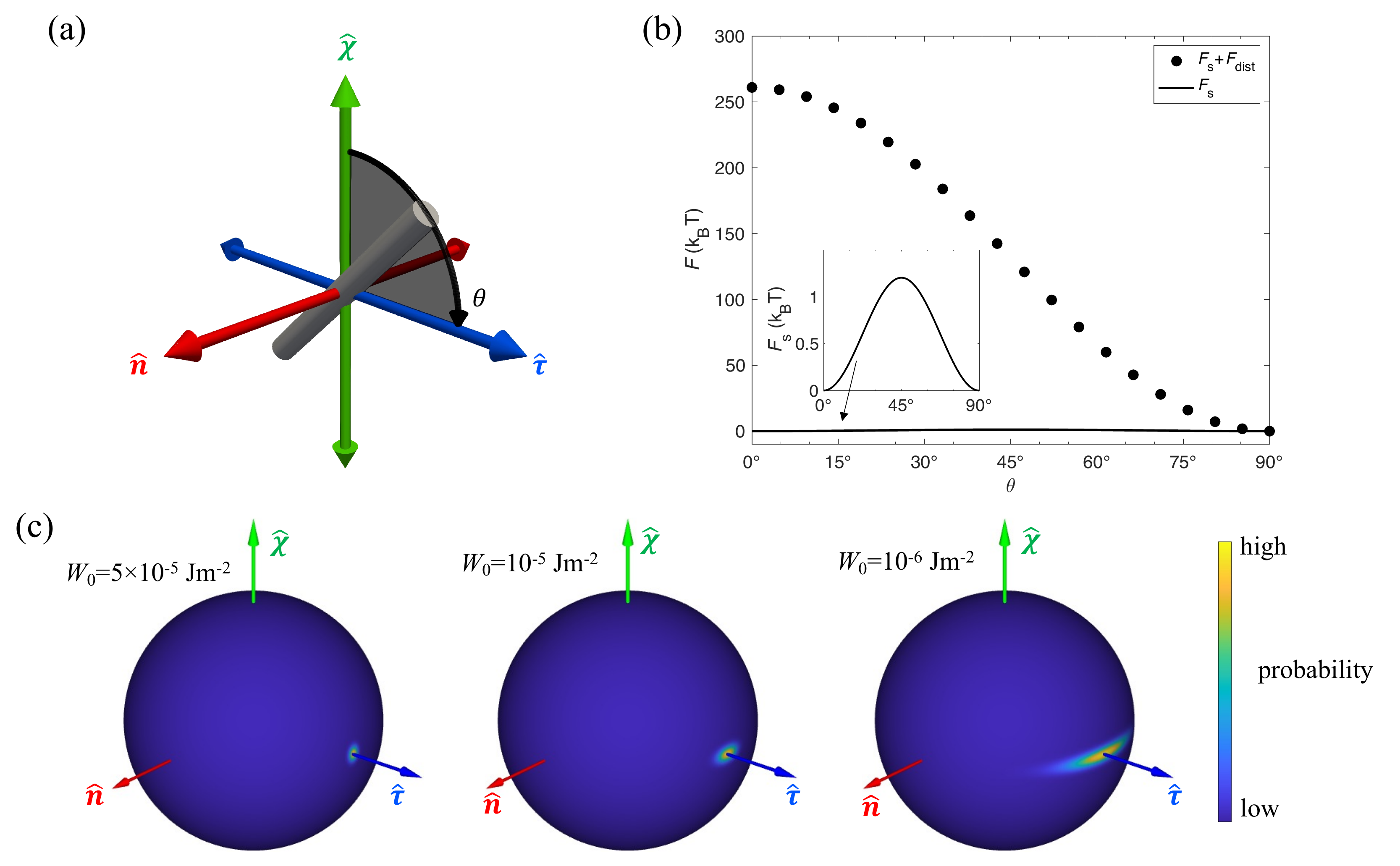}
    \caption{(a) The orientational fluctuation of a cylindrical rod (grey) along angle $\theta$ in the molecular frame. (b) Corresponding energy values are calculated based on \eq{fstot} with and without the LC elastic distortion energy. The inset shows that the surface energy is overpowered by the elastic counterpart. (c) Unit sphere projection of the orientational distribution of thin rods dispersed in a chiral LC. Surface anchoring strength $W_0=10^{-6} \rm{J m^{-2}}$ for (a)-(b) and cholesteric pitch $p=30 {\rm \um}$, $\Lc=1.7\um$ and $\Dc=28 \rm{nm}$ are used for all calculations.}
    \label{theory_rod_symbreak}
\end{figure*}

Gathering the findings of the previous paragraph we revisit the realigning potential acting on a rod immersed in a cholesteric host. The total external potential is given by the bare Rapini-Papoular contribution \eq{usurf} for the undistorted host director plus the free energy contributions from elastic distortions:
\beq
F_{s, {\rm tot}}  \sim  F_{s}  + \Delta F_{\rm dist} 
\label{fstot}
\eeq
Since the distortion term cannot be resolved for any rod orientation but only for cases when the rod is aligned along either of the directions of the local frame ($\bn, \btau, \bchi$) of the helical LC host frame we use the following interpolation form:
\begin{align}
\Delta F_{\rm dist} (\eta, \gamma)  \sim & \Delta F_{\rm twist} \sin^{2} \eta  + \Delta F_{\rm tilt} \cos^{2} \eta \sin^{2} \gamma  
\label{frodinterp}
\end{align}
in terms of the two  angles $\eta = \theta - \tfrac{\pi}{2}$ and $\gamma = \delta - \tfrac{\pi}{2}$ represented in \fig{exp_homeo_rod}(c,f) and key elastic  contributions;  $\Delta F_{\rm tilt} = F(\bhu \parallel \bn ) - F( \bhu \parallel \btau )$ associated with tilting the rod away from the $\tau$-axis towards the $\bn$-direction, discussed in Appendix A,  and $\Delta F_{\rm twist} = F(\bhu \parallel \bchi ) - F( \bhu \parallel \btau ) $ [\eq{twidi}] the energy cost associated with twisting the surface disclination wrapped along the body of the cylinder. From the analysis in the previous section, we found that $\Delta F_{\rm twist}$ is a few hundreds of $k_{B}T$ [\fig{theory_rod_symbreak}(b)] whereas the elastic distortions due to tilting are much weaker ($\Delta F_{\rm tilt} < k_{B}T$) and may, in fact, be neglected altogether for the weak anchoring regime considered in this study (Appendix A).  The  elastic energy is then minimal (zero) when the rods align along the $\btau$ directions ($\theta^{\ast} = \pi/2$ and $ \delta^{\ast} = \pi/2 $) as observed in our experiments [\fig{exp_homeo_rod}]. The best correspondence with experimental data is obtained for a surface anchoring amplitude of about $W_{0} \sim 6 \times 10^{-7} {\rm J m^{-2}}$. An overview of the orientational probability distributions associated with \eq{fstot}, based on the Boltzmann exponent \eq{fsingle},  are depicted in \fig{theory_rod_symbreak}(c) indicating that the rod preferentially aligns along the $\btau$-axis with considerable orientational biaxiality developing around the main alignment direction.

For the case of homeotropic rods reported in \fig{exp_homeo_rod} we may roughly estimate the energy contribution due to the twisted disclination from the width of the distributions depicted in panels (c) and (f). For small angles $\eta$ the Boltzmann factor of \eq{frodinterp} translates  into a simple Gaussian distribution:
\beq
f(\eta) \propto \exp (- \Delta F_{\rm twist} \eta^{2} )
\eeq
 and we identify a standard Gaussian ${\rm FWHM} =2.355/\sqrt{2 \Delta F_{\rm twist}}$. This subsequently gives $\Delta F_{\rm twist} \approx 22 k_{B}T$ for homeotropic rods with  $L_{\rm{c}} = 1.7 \um $  and $\Delta F_{\rm twist} \approx 76 k_{B}T$  for the longer rods  with $L_{\rm{c}} = 3 \um $ suggesting that, in both cases, the thermal motion of the rods is assuredly insufficient to overcome the energy barrier between the $\tau$ and $\chi$ alignment directions. The values are in qualitative agreement with the prediction from our analytical model \eq{twidi} where $\Delta F_{\rm twist} \propto L_{\rm{c}}^{3}$ suggests that the elastic energy cost of orientating the rods from $\btau$ to $\bchi$ directions is indeed quite sensitive to the colloidal rod length $L_{\rm{c}}$. 
 The actual values from  \eq{twidi}, however, should be considered as an upper bound for $\Delta F_{\rm twist}$ mainly because in our model the local nematic order parameter $\Sm$  of the host is constrained at its far-field bulk value and is not allowed to relax in regions where director distortions are the largest, as observed in our experiment and simulations.

\subsection{Colloidal order parameters}

In order to facilitate comparison with experimental results, we define the colloidal orientational order  which measures the principal direction of alignment of the colloids along the cholesteric helix. 
Taking the local molecular LC director $\bn$ as a reference frame we define a colloidal uniaxial order parameter as follows:
\beq
S_{\rm cm} = \langle {\mathcal P}_{2} (\bhu \cdot \bn)\rangle_{f} 
\label{s2}
\eeq
with $\avg{...}_f$ denoting a thermal average, and a colloidal biaxial nematic order parameter that measures the relative orientational order with respect to the principal directions orthogonal to $\bn_{h}$: 
\beq
\Delta_{\rm cm} = \langle (\bhu \cdot \btau)^{2} - (\bhu \cdot \bchi)^{2} \rangle_{f}  
\label{d2}
\eeq
Alternatively, we can probe the orientational order from the tensorial order parameter for colloids ${\bf Q}_{\rm{c}} = \frac{3}{2} \langle \bhu \otimes \bhu  \rangle_{f} - \frac{1}{2}\bf{I} $ which measures orientational order with respect to the principal colloidal alignment direction independently from the chosen reference frame. The corresponding uniaxial and biaxial order parameters defined within the colloidal frame are denoted by $S_{\rm cc}$ and $\Delta_{\rm cc}$, respectively.
In case of colloids aligning along the molecular director $\bn$, the two frames coincide and the corresponding values of order parameters are identical.

\hide{
\begin{figure*}
	\includegraphics[width=2\columnwidth]{OPs.pdf}
	\caption{(a) Biaxial and uniaxial order parameters of colloidal rods with various surface anchoring strengths $W_0$ calculated using $L=1.7 \um, p=30 \um$ (solid lines) and $L=3 \um, p=150 \um$ (dashed lines). Filled and empty circles correspond to experimental results \fig{exp_homeo_rod}(a) and (d), respectively. (b) Colloidal order parameters of thin rods obtained with different values of LC chirality, with theoretical values (solid and dashed lines) respectively matching those measured experimentally (filled and empty circle). }
	\label{OPs}
\end{figure*}
}

\begin{table}[h!]
\centering
    \begin{tabular}{c  c  c  c  c} 
    \hline\hline
    Sample & $S_{\rm{cc}}$ & $\Delta_{\rm{cc}}$ & $S_{\rm{cm}}$ & $\Delta_{\rm{cm}}$ \\  
    \hline
    Homeotropic  disk & 0.66 & 0.067& 0.66 & 0.067 \\ 
    
    Planar rod & 0.94 & 0.014 & 0.94 & 0.014 \\
    
    Homeotropic rod ($L_{\rm{c}}=1.7 \um$) & 0.70 & 0.12 & -0.26 & 0.76 \\
    
    Homeotropic rod ($L_{\rm{c}}=3.0 \um$) & 0.86 & 0.065 & -0.38 & 0.89 \\
    \hline\hline
    \end{tabular}
    \caption{Colloidal order parameters measured in colloidal coordinates (c) and molecular frame (m) for each set of experiments shown in \fig{exp_homeo_disk}, \fig{exp_planar_rod}, and \fig{exp_homeo_rod}.}
    \label{table_OPs}
\end{table}
To quantify the symmetry-breaking of the colloidal orientational distribution in experiments, we measured the uniaxial $S_{\rm{cc}}$ and biaxial $\Delta_{\rm{cc}}$ order parameters for both disks and rods (Table ~\ref{table_OPs}). The uniaxial order parameter $S_{\rm{cc}}$, as a measure of unidirectional ordering (\eq{s2}), represents the strength of orientational confinement which greatly depends on the synthesized materials and the ensuing surface anchoring effects. Subsequently, the non-equivalence of axes orthogonal to the average colloidal/molecular axis is evaluated using the biaxial order parameter $\Delta_{\rm{cc}}$ (\eq{d2}), with $-1<\Delta_{\rm{cc}}<1$.
The values of $S$ are experimentally determined to be $S_{\rm{cc}}=S_{\rm{cm}}=0.66$ for homeotropic disks dispersed in chiral 5CB-based LC [\fig{exp_homeo_disk}] and $S_{\rm{cc}}=S_{\rm{cm}}=0.94$ for rods with planar boundary condition [\fig{exp_planar_rod}]. The values of $\Delta_{\rm{cc}}$ are found to be 0.067 and 0.014, respectively, showing a robust symmetry-breaking among $\bchi$ and $\btau$ leading to biaxial orientational symmetry.

When the average orientations of the two components differ, however, the choice of reference frame determines the values of order parameters. Stronger orientational fluctuations are found for the shorter rods with homeotropic anchoring (\fig{exp_homeo_rod} a-c) with $S_{\rm{cc}}=0.70$ and $\Delta_{\rm{cc}}=0.12$, while the dispersion of the longer rods showed a smaller orientational distribution with higher uniaxial order parameter $S_{\rm{cc}}=0.86$ and lower biaxiality $\Delta_{\rm{cc}}=0.065$, though still higher than that measured for planar rods. 
The order parameters obtained are in good agreement with the analytical prediction using \eq{ht_fluc_new} and \eq{frodinterp}, which give $S_{\rm{cc}}=0.75, \Delta_{\rm{cc}}=0.17$ for the shorter rods, and $S_{\rm{cc}}=0.90,\Delta_{\rm{cc}}=0.058$ for the longer rods using the experimental parameters.

The enhanced biaxiality for homeotropic rods aligning perpendicular to the molecular director $\bn$ will be discussed in the following section.
Calculated in the molecular reference frame, the negative values of $S_{\rm{cm}}$ and large values of $\Delta_{\rm{cm}}$ simply represent the geometry in which the average colloidal director lies orthogonal to the molecular one $\bn$.

\hide{ 

At infinitely low rod concentrations, the order parameters $S$ and $\Delta$ defined within the frame of the  molecular host (\eq{s2} and \eq{d2}) are readily computed from the Boltzmann factor 
  $f( \bhu ) = \mathcal{N} \exp ( - \beta  F_{s, {\rm tot}} (\delta, \theta)   ) $.
An overview of the results as a function of the anchoring strength $\bar{W} = \beta W_{0} LD$ is given in \fig{ww}. The best correspondence with experimental data ($S \sim -0.33 -0.43$ and $\Delta = 0.6 - 0.8$) is found for $ \bar{W} \sim 6 k_{B}T$ which corresponds to a surface anchoring  amplitude of about $W_{0} \sim 6 \times 10^{-7} {\rm J m^{-2}}$.

}

\section{Discussion}

\subsection{Enhanced biaxial symmetry-breaking at perpendicular colloidal-molecular alignment}
For colloidal rods immersed in a chiral LC, the biaxial order developed at the level of the colloids is much more pronounced for rods with homeotropic boundary condition, whose energy-favored orientation is along $\btau$ and perpendicular to $\bn$. As a consequence, the rotational symmetry (with the rotation axis being $\btau$) is obviously not continuous, with $\bn$ being the material axis representing the actual molecular direction and $\bchi$, in contrast, an ``imaginary" one. The dissimilarity and the resulting symmetry breaking are thus more pronounced than those in the case of colloids aligned along $\bn$, as clearly shown above by biaxial colloidal distribution probabilities in the experimental results.

Analytical theory predicts a similar type of realignment and enhanced biaxiality to occur for homeotropic rods immersed in chiral host LCs where the chiral and biaxial ``dressing" around the rod will also be much more pronounced than that of rods with planar or parallel boundary conditions.
With the significant contribution from the elastic energy of the background molecular LC to the total free energy [\eq{frodinterp}], we are provided an additional control of this emergent biaxiality by tuning elasticities of the molecular host to boost the biaxiality of the hybrid LC [\eq{Ftwistel}].
Likewise, disks with planar anchoring, which could be realized through appropriate surface functionalization \cite{mundoor2019electrostatically}, exhibit equivalent perpendicular alignment (but this time along the helical axis $\bchi$) which would also give strongly enhanced biaxial order in the disk orientation distribution.

\subsection{Quadratic scaling of biaxial order parameter with chirality}
\begin{figure*}
	\includegraphics[width=2\columnwidth]{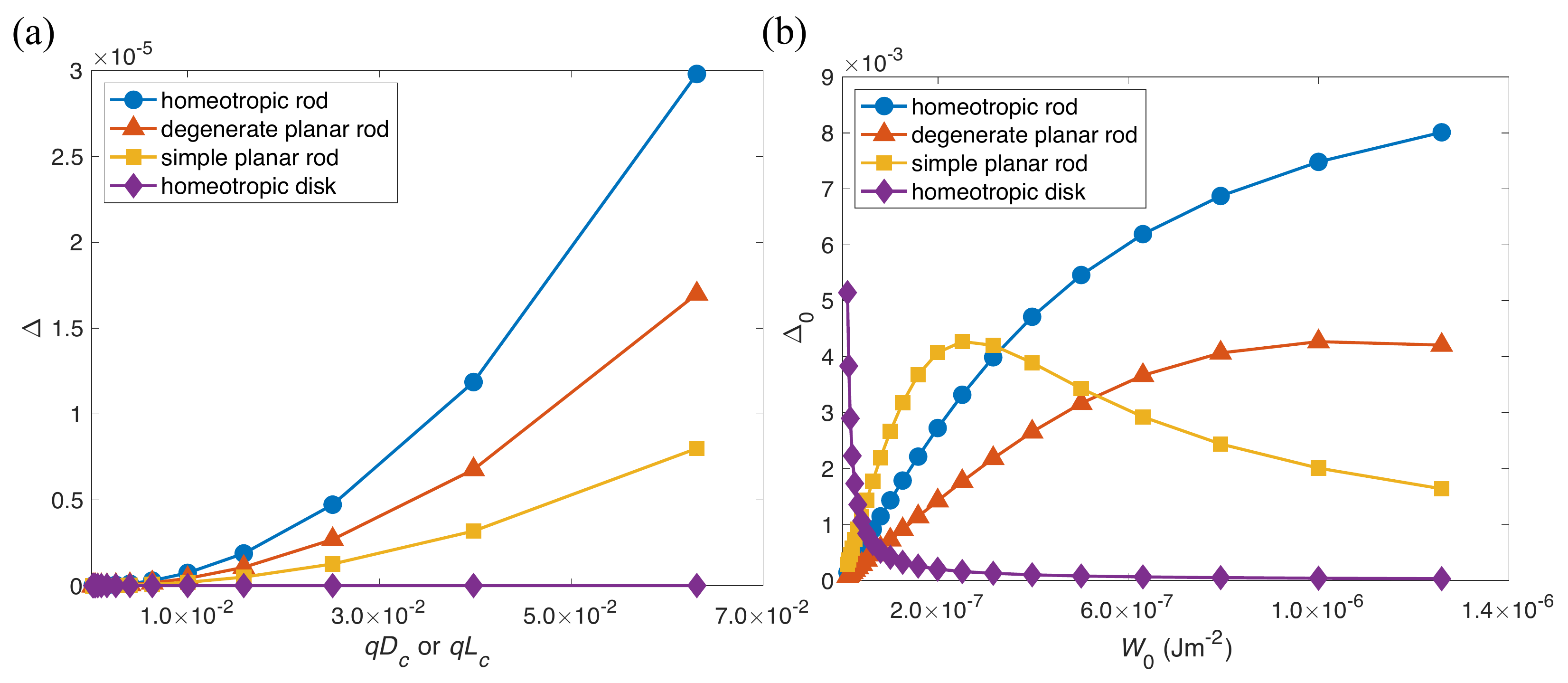}
	\caption{(a) Biaxial order in the colloidal orientation distribution for the weak chirality regime $qa \ll 1$, with $a$ the typical colloid size. The results are based on the Rapini-Papoular surface anchoring energy \eq{rapo} and \eq{fsingle} using $W_{0} = 10^{-6} \rm{J m^{-2}}$. (b) Dependence of the prefactor $\Delta_{0}$, defined by $\Delta \sim \Delta_{0}(qa)^2$ on the anchoring strength based on the colloidal dimensions $L_{\rm c}=1.7 {\rm \um}$ and $D_{\rm c}=28 \rm{nm}$ for the rods and $D_{\rm c}=2 {\rm \um}$ for the disks. }
	\label{theory_biaxiality_scaling}
\end{figure*}

In the weak molecular chirality regime, we may characterize the leading order contribution of chirality to colloidal biaxiality by expanding the biaxial order parameter up to the quadratic order in the inverse pitch $q=2\pi/p$:
\beq
\Delta=\Delta_0(qa)^2+\mathcal{O}[(qa)^4]
\eeq
where the length scale corresponds to the colloidal dimensions; $a=D_{\rm c}$ for thin disks and $a=L_{\rm c}$ for cylinderical rods.
The zeroth order term must be zero given that no intrinsic biaxiality can be expected from purely uniaxial components at zero chirality. Also, the linear term proportional to $qa$ must vanish since the value of biaxiality should not depend on the handedness of the host material.
Following the results \eq{fsingle}, \eq{d2}, and the free energies for each type of colloids, we computationally verify the quadratic scaling $\Delta \sim \Delta_0 (qa)^2$ within the weak chirality approximation $qa \ll 1$ [\fig{theory_biaxiality_scaling}(a)].
Interestingly, the quadratic scaling of biaxial order with the parameters $qL_{\rm c}$ or $qD_{\rm c}$ resembles the theoretical prediction by Priest and Lubensky for a single-component molecular LC \cite{priest1974biaxial}, in which case $L_{\rm c}$ needs to be replaced by $L_{\rm m}$ denoting the size of the molecules. Despite the different derivations of biaxiality from component material(s), the agreement between the results from our hybrid LC system and single-compound LC reveals the underlying physical principle, namely a close relationship between biaxial order and chirality. Most interestingly, the prefactor $\Delta_0$ turns out to be very different for each system considered and has a distinct, non-trivial dependence on the surface anchoring strength [\fig{theory_biaxiality_scaling}(b)].

\subsection{Enhanced biaxiality of the molecular host at the colloidal surface}

The molecular biaxial order parameter $\Delta_{\rm{m}}$ measures the broken uniaxial symmetry of the LC host (which is 5CB).  
The $\Delta_{\rm{m}}$ is associated to the tensorial local mean-field order parameter by \cite{mundoor2021,mottram2014introduction}:
\beq
    \bQ = \Sm \left(\frac{3}{2} \bn \otimes \bn - \frac{\bf{I}}{2} \right) + \Delta_{\rm{m}} \left( \frac{3}{2}  {\bhatm \otimes \bhatm} - \frac{\bf{I}}{2} \right)
    \label{Qmdef}
\eeq
with the molecular director field $\bn$ and the biaxial director $\bhatm$ orthogonal to each other. 
Here, $\Sm$ is the scalar order parameter measuring the uni-directionality of $\bn$, with $\Sm \geq \Delta_{\rm{m}} \geq 0$.
Accordingly, in the numerical computation, the order parameters are determined by the diagonalization of the Q tensor: 
\begin{align}
    \Delta_{\rm{m}} &= \frac{2}{3} (\lambda_2-\lambda_3) \nonumber \\   
    \Sm &= \lambda_1+\Delta_{\rm{m}}/2
    \label{mOPs}
\end{align}
where $\lambda_1>\lambda_2>\lambda_3$ are the eigenvalues of $\bQ$. The directors $\bn$ and $\bhatm$ are then found by calculating the eigenvectors corresponding to $\lambda_1$ and $\lambda_2$, respectively. 
Since eigenvalues are interpreted as the ``directionalities'' along each orientation (eigenvector), the calculation of $\Delta_{\rm{m}}$ in \eq{mOPs}  corresponds exactly to finding the inequivalence of the two minor axes ($\bhatm$ and $\bn \times \bhatm$), and the value of biaxiality is a measure of the broken rotational symmetry along $\bn$, analogous to the colloidal orientation distributions illustrated above.
Using numerical modeling based on the Q-tensor representation of the LC order parameters, we find $\Delta_{\rm{m}}$ at a far-field helical background to be of the order of $10^{-7}$, which is precisely the value predicted using $\Delta_{\rm{m}} \sim (qL_{\rm{m}})^2$ with the size of a 5CB molecule being in the nanometer range $L_{\rm{m}}=2 \rm{nm}$ \cite{priest1974biaxial}, showing the intrinsic biaxial order in the molecular chiral liquid crystal.

\begin{figure}
	\includegraphics[width=1\columnwidth]{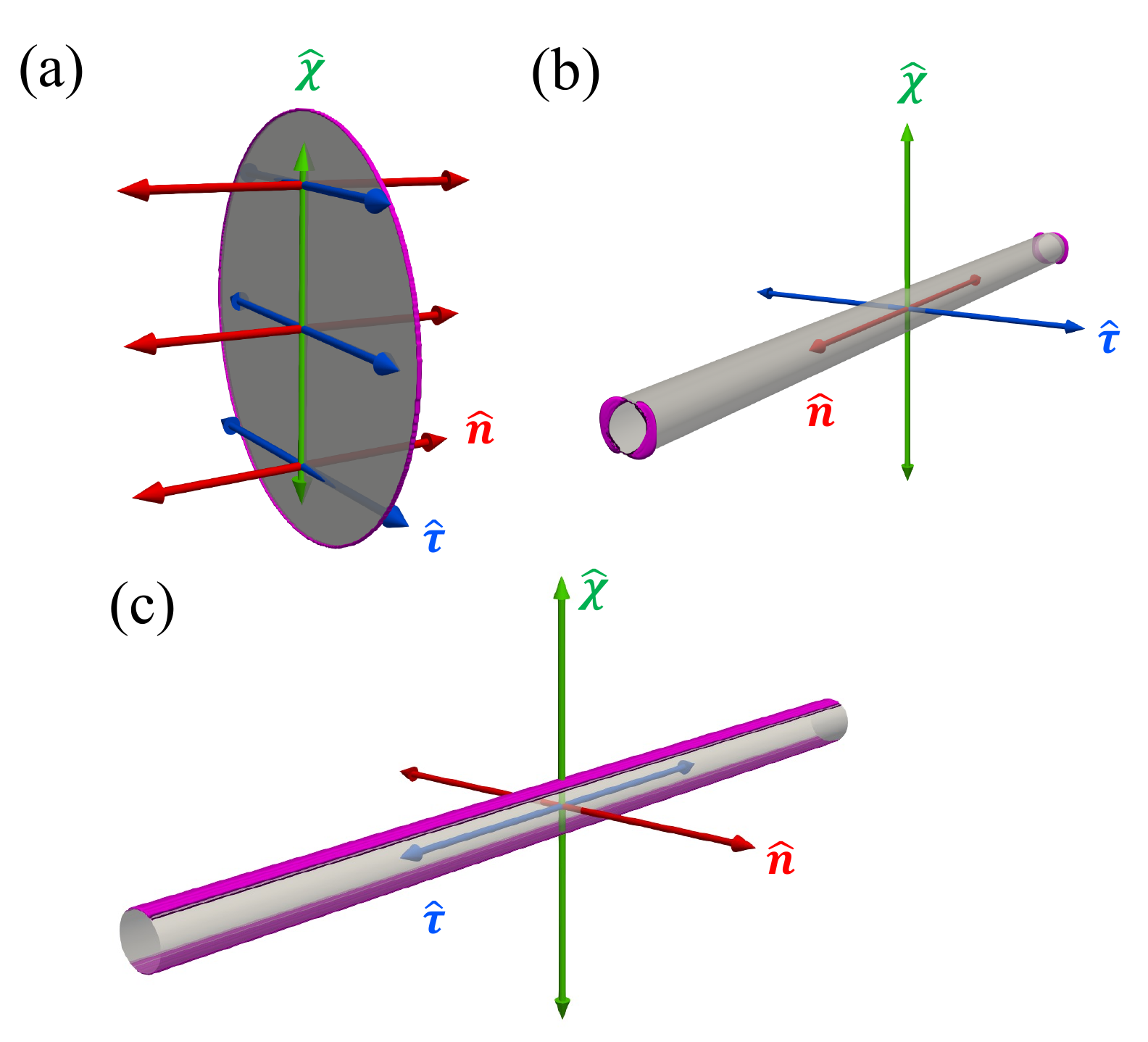}
	\caption{(a)-(c) Contours of molecular biaxiality (magenta) around the colloids (gray) marking the regions with $\Delta_{\rm{m}}$ larger than $10^{-3}$ (a) and $10^{-4}$ (b,c), respectively. The orthogonal frame defining the molecular axes is colored as in \fig{chiral_schematic}. Homeotropic anchoring condition is used for (a,c) and planar anchoring for (b). Surface anchoring strength $W_{0}=10^{-6} {\rm J m^{-2}}$ and LC helical pitch $p=30 {\rm \um}$ is used for all simulations.}
	\label{molecular_biaxiality_contour}
\end{figure}

Interestingly, we also discover that $\Delta_{\rm{m}}$ greatly increases from $10^{-7}$ in the far-field limit to $10^{-4}$ or even $10^{-3}$ near the colloidal surfaces [\fig{molecular_biaxiality_contour}], being especially prominent at the regions where the surface anchoring force favors a distinct molecular director alignment from the helical far-field. The enhanced biaxiality induced by the colloidal particles is qualitatively interpreted as the mismatch of two axes -- the particle surface anchoring orientation $\bv$ and background LC aligning direction $\bn$. 
As a quantification of the broken uniaxial rotation symmetry of $\bn$, higher values of $\Delta_{\rm{m}}$ can be found at particle surfaces with a greater discrepancy in the two orientations, with maximum $\Delta_{\rm{m}}$ located at regions where surface normal director perpendicular to background far-field [\fig{molecular_biaxiality_contour}]. 
Furthermore, within LC regions with a $\Delta_{\rm{m}}$ dominated by particle surface and far exceeding the background value $10^{-7}$, the biaxial director $\bhatm$ is found to coincide with the perpendicular component of the surface anchoring director to the nematic director $\bhatm = \bv-(\bv \cdot \bn) \bn$, confirming the idea that the colloidal surface induces  molecular biaxiality order by introducing an energy landscape for $\bn$ without uniaxial symmetry.

In case of no host chirality, biaxial order stabilized by correlations between colloidal particles immersed in nematic 5CB was reported in \cite{mundoor2018}.
Furthermore, compared to pure molecular LCs without colloids, the frustrated alignment of $\bn$ induced by the presence of colloidal particles also leads to a reduced bulk nematic order parameter $\Sm$ and the formation of defects in cases with strong surface anchoring.
In our systems, though, we expect the two independent contributions to the ``biaxialization'' of uniaxial 5CB liquid crystal -- the introduction of chiral dopant and of colloidal particles -- to have negligible effects on the free energies in our analytical model, which is evident by the induced values of $\Delta_{\rm{m}}$ and has been confirmed by the numerical modeling using a tensorial order parameter $\bQ$.

\hide{
Both effects change the ground-state arrangement of liquid crystal molecules and impose energy landscapes without continuous rotation symmetry, which is disturbed by the magnitude and orientation of twisting alignment and particle surface anchoring, respectively.
}

\subsection{Biaxial interpretation of chiral liquid crystals}
\begin{figure}
	\includegraphics[width=1\columnwidth]{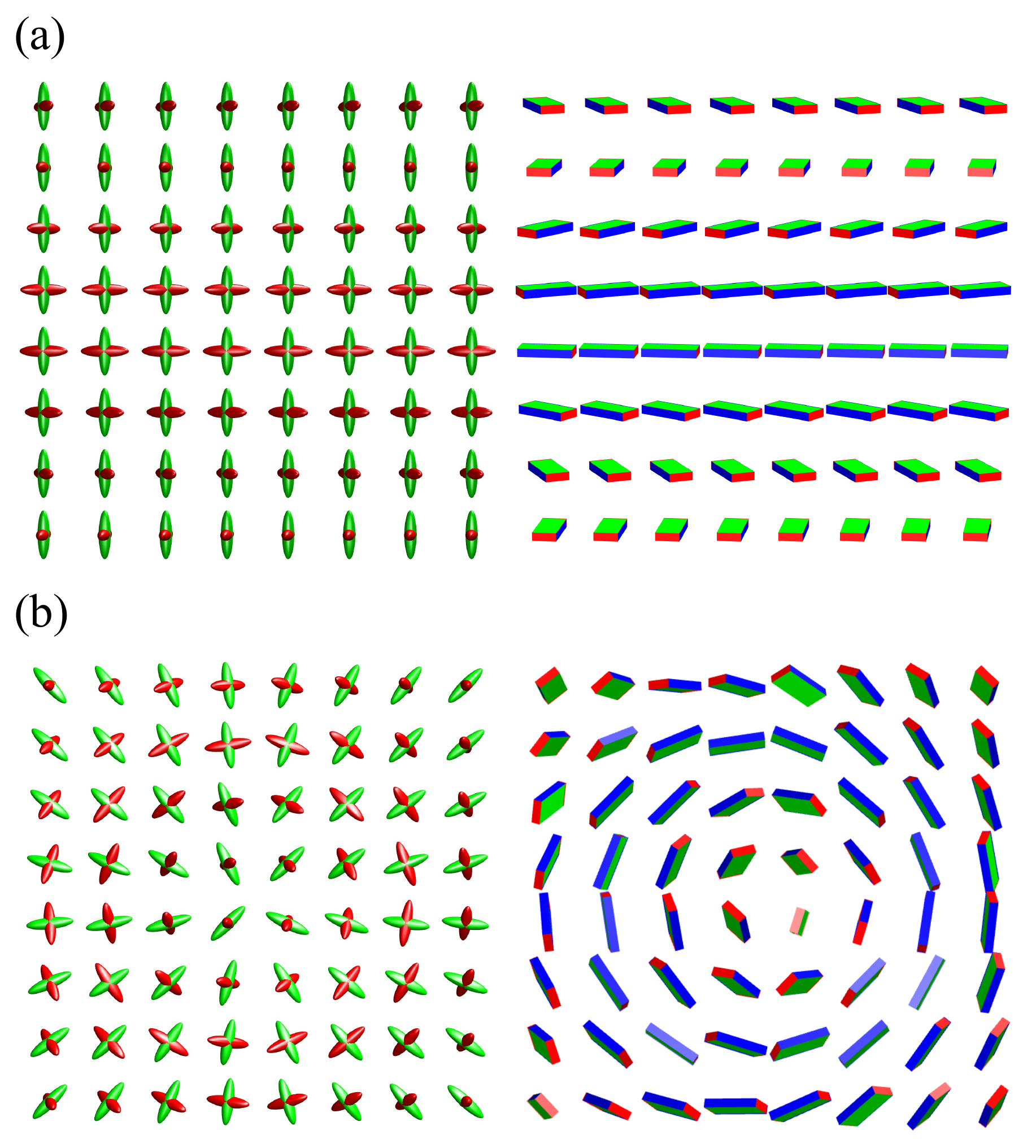}
	\caption{(a) The director profiles simulated inside a Bloch-wall-like structure resembling a helical-twist. Treated as in a uniaxial LC, $\bn$ (red) and $\bchi$ (green) are calculated using chirality tensor \cite{beller2014geometry} and visualized as ellipsoids (left). The directors simulated instead by biaxial Q-tensor are visualized using bricks (right), with red, blue, and green faces respectively corresponding to principle $\bn$, biaxial $\bhatm$, and the third $\bn \times \bhatm$ orthogonal axes \cite{mottram2014introduction}. (b) Numerical simulation of molecular $\bn$ and helical $\bchi$ axes in a 2D meron-like structure using chirality-based (left) and biaxiality-based approaches (right).}
	\label{chiral-vs-biaxial}
\end{figure}

As suggested in the section above, the intrinsic biaxiality of a chiral nematic LC allows us to define local biaxial directors even in absence of colloidal particles. The molecular biaxial order persists $\Delta_{\rm{m}} \sim (qL_{\rm{m}})^2$ as long as the chirality $q$, or the helicity in the director alignment, is non-vanishing. To accurately account for this unavoidable biaxiality, we modify and expand the calculation in Ref. \cite{efrati2014orientation,beller2014geometry} for uniaxial chiral nematics, in which the chirality-associated directors ($\bn$,$\bchi$,$\btau$) are found by diagonalizing a 3-by-3 handedness tensor $\bf{H}$ defined as:
\beq
    {\bf{H}}_{ij}={\epsilon}_{ikl} \bn_{k} \pdn{l}{j}
    \label{htuniaxial}
\eeq
with summation over indices assumed. The trace $\sum_i {\bf H}_{ii} =  -\bn \cdot (\nabla \times \bn )$ gives the helicity of the LC director alignment field. Considering the intrinsic biaxial order in chiral LCs, we can similarly construct the handedness tensor using the molecular tensorial order parameter $\bQ$:
\beq
    {\bf{H}}_{ij}=\frac{4}{9{\Sm}^2} {\epsilon}_{ikl} \bQ_{kn} \pdQm{ln}{j}
\eeq
The uniaxial definition \eq{htuniaxial} can be recovered by expanding the equation using \eq{Qmdef} with $\Delta_{\rm{m}}=0$. Note that the trace of the handedness tensor again represents the helicity and is identical to the chiral part of elastic free energy ($L_4$ term in \eq{LdGbulk}).
Strikingly, in numerical simulation we discovered that the helical director field $\bchi$, which is computed as the eigenvector corresponding to the eigenvalue with the largest absolute value, thoroughly matches the directors calculated by diagonalizing $\bQ$: $\bchi=\bn \times \bhatm$ [\fig{chiral-vs-biaxial}], which also immediately suggests $\btau=\bhatm$ (Note all directors are head-tail symmetric). 
The excellent overlap of the two orthogonal frames, ($\bn$, $\btau$, $\bchi$) originating from chirality and ($\bn$, $\bhatm$, $\bn \times \bhatm$) representing biaxiality, directly demonstrates the biaxial feature in chiral nematic LCs.
The energy-minimizing of $\bQ$ automatically incorporates these symmetries once all degrees of freedom beyond those for pure uniaxial nematics are allowed.
Consequently, one can straightforwardly identify  chirality through the concomitant biaxial properties using $q \sim \sqrt{\Delta_{\rm{m}}}/a $ and $\bchi=\bn \times \bhatm$ instead of investigating the helical twisting and spatial derivatives of the LC directors. These values are well-defined from the biaxiality calculation even inside LC defects with reduced uniaxial order parameter $\Sm$.

Therefore, with the chirality-driven biaxial symmetry taken into account, one can naturally analyze structures within a chiral LC using considerations similar to the ones derived for biaxial nematics [\fig{chiral-vs-biaxial}].
Since the theory describing the topological classification of defects and solitons in biaxial nematics, which has an order parameter space $ SO(3)/D_2 $, is distinct from those emerging in a uniaxial LC with $ \mathbb{S}^2 / \mathbb{Z}_2 $ counterpart
\cite{mermin1979topological,wu2022hopfions,luckhurst2015biaxial,gennes-prost,lavrentovich2001cholesteric}, the biaxial symmetry in a chiral LC offers an alternative interpretation of topological objects in cholesterics differing from their more conventional description.
For example, a helical configuration resembling a Bloch wall can be found across our experiments. By identifying the $\bchi$ director field within, which is uniformly aligned, we can visualize the configuration as a 1-dimensional soliton formed in brick-like LCs with matching director fields [\fig{chiral-vs-biaxial}] with a uniform $\bchi$ field and helical twisting in $\bn$ and $\btau$ fields.
Furthermore, unlike uniaxial LCs with a single director field, biaxial systems with three orthogonal director fields cannot accommodate a fully nonsingular solitonic structure as 2D translationally invariant fully nonsingular objects, implied by $\pi_2 (SO(3)/D_2) = 0$ \cite{wu2022hopfions,mermin1979topological}. 
As shown in \fig{chiral-vs-biaxial}(b), a meron-like arrangement of directors is a nonsingular soliton embedded in the molecular director $\bn$. 
The meron, or half-skyrmion structure, has been constructed as a 2D soliton composed of a single director or vector with the absence of singularity \cite{skyrme1962unified,leonov2014theory,tai2020surface}.
The structure becomes, however, a singular defect in a biaxial system, as demonstrated by the emergence of  singularities found at the center in the $\bchi$ and $\btau$ director fields orthogonal to the material director field.
Similarly, 3D topological solitons, fingers, and other non-singular structures in cholesterics can be viewed as defects lines and loops in a biaxial system by thoroughly analyzing all three directors as well \cite{wu2022hopfions,machon2016umbilic}.
Besides, some phenomena of the defects and soliton structures in a system of chiral nematics, including nonabelian disclinations and their entanglement behaviors, are elucidated only from the perspective of biaxial topological descriptions that are distinct from uniaxial topology \cite{wu2022hopfions,lavrentovich2001cholesteric,kurik1988defects,alexander2012colloquium}.
With the biaxial directors defined and simulated in consistency with the chiral description, the biaxial features of chiral nematics, including their topological defects, solitons, and frustrated structures can be easily and naturally explored. This opens up the possibility of using molecular-colloidal chiral nematics as model systems in the exploration of nonabelian vortices, solitonic structures with low-symmetry order parameters, etcetera.

\section{Conclusion and outlook}

\begin{figure*}
	\includegraphics[width=1.8\columnwidth]{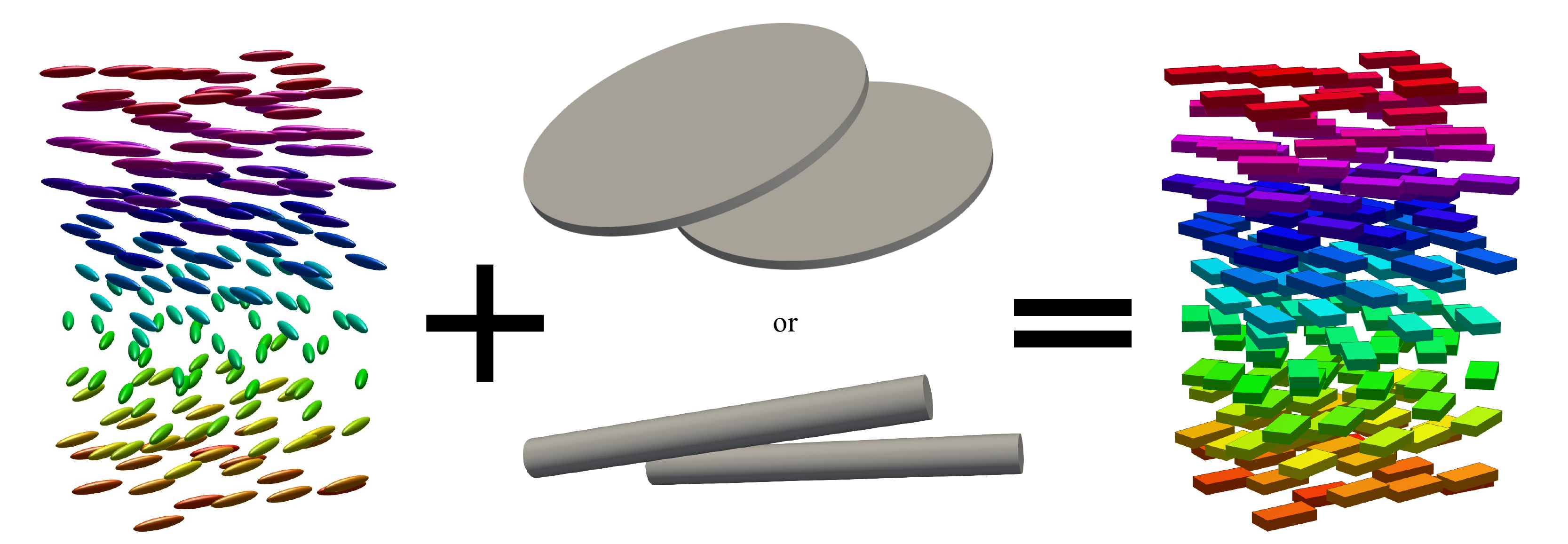}
	\caption{Our molecular-colloidal hybrid system with emergent biaxial symmetry consists of purely uniaxial building blocks. The chirality effects at different scales yield effective behavior of a biaxial chiral molecular-colloidal LC. }
	\label{biaxial_schematic}
\end{figure*}

We have explicitly demonstrated that immersing uniaxial, non-chiral colloidal rods and disks into a low-molecular-weight cholesteric liquid crystal host leads to emergent biaxial order that we identify at both colloidal and molecular levels by combining experiment with numerical simulation and analytical theory [\fig{biaxial_schematic}]. Unlike the previously studied case of hybrid molecular-colloidal biaxial phases \cite{liu2016,mundoor2021,mundoor2018}, we observe multi-level biaxial symmetry-breaking at ultralow colloidal content where colloid-colloid interactions are negligible. By exploring a variety of colloidal shapes and surface anchoring symmetries we report biaxial order emerging at three distinct levels. First, molecular director distortions develop around each colloid which, although being of marginal extent because of weak surface anchoring conditions, display a distinct two-fold signature imparted by the cholesteric host. Second, the orientational distribution of the colloids around the local cholesteric director is demonstrated to adopt a clear biaxial signature, and the response of the corresponding biaxial order parameter is found to depend non-trivially upon the surface anchoring strength as well as on the ratio of the cholesteric pitch and the principal colloidal dimension (rod length or disk diameter). Finally, at the molecular scale, we demonstrate that enhanced biaxiality emerges close to the colloidal surface at levels strongly exceeding those expected for purely molecular cholesterics.

A particularly striking manifestation of biaxial symmetry-breaking is encountered for thermotropic cholesterics doped with colloidal rods with homeotropic surface anchoring. Driven by a combination of surface anchoring forces and an energy penalty incurred by twisting a weakly developed surface disclination along the rod main axis, these rods have a strong tendency to align perpendicular to both the helical axis and the local cholesteric director, thus imparting a two-fold $D_{2h}$ orientational symmetry onto the hybrid system at each point along the cholesteric helix.
By means of numerical minimization of the Landau de Gennes energy and mean-field theory based on the Rapini-Papoular surface anchoring energy, we have revealed that the multi-level expression of emergent biaxiality in our systems  is already manifest  at ultralow colloid concentrations, essentially as a single-colloid effect, and we find consistent agreement between our predictions from modeling and the experimental observations.

Our results pave the way towards controlled biaxial order at both colloidal and molecular levels. By harnessing the interplay of chiral and biaxial symmetries, future research efforts could be directed along the following several emergent avenues.
At larger colloidal concentrations a richer phenomenology could be expected and explored due to the more prominent roles expected to be played by steric, electrostatic or defect-mediated colloid-colloid interactions further enriching the surface anchoring and elastic forces discussed here. Besides the emergent symmetry breaking discussed here, one could, in principle, also apply electric or magnetic fields to reconfigure either molecular or colloidal sub-systems, or both, to achieve even lower externally induced symmetries of LCs, for instance, corresponding to  triclinic or monoclinic point groups.
Finally, by realizing topological solitons in the molecular-colloidal hybrid system with nontrivial chirality and biaxiality, one could reveal the stability of topological structures for various low-symmetry order parameter spaces. While ferromagnetic colloidal particle dispersions have already provided insight into the possibility of formation of solitons in polar chiral liquid crystals \cite{ZhangPRL}, this study could be extended to symmetries differing from nonpolar and polar uniaxial LCs, for example, by exploring multi-dimmensional solitonic structures corresponding to the $ SO(3)/D_2 $ order parameter space.

\section*{ACKNOWLEDGMENTS}

We acknowledge discussions with M. Bowick, T. Lee, T.C. Lubensky, B. Senyuk, M. Ravnik and M. Tasinkevych. We are grateful to T.C. Lubensky for providing helpful suggestions and feedback on the initial versions of this manuscript. The experimental and numerical simulations research at University of Colorado Boulder was supported by the US Department of Energy, Office of Basic Energy Sciences, Division of Materials Sciences and Engineering, under contract DE-SC0019293 with the University of Colorado at Boulder. M.T.L. and H.H.W. acknowledge financial support from the French National Research Agency (ANR) under grant ANR-19-CE30-0024 ``ViroLego". 
I.I.S. acknowledges the support of the International Institute for Sustainability with Knotted Chiral Meta Matter at Hiroshima University in Japan during part of his sabbatical stay, as well as the hospitality of the Kavli Institute for Theoretical Physics in Santa Barbara, when he was partially working on this manuscript. 
This research was also supported in part by the National Science Foundation under Grant No. NSF PHY-1748958 (I.I.S. and J.-S.W.).


 \section*{Appendix A: Elastic distortions around the rod surface for $\bhu \perp \bchi$ }

In order to complete our understanding of the strength of the elastic distortions surrounding the main section of a thin rod we now focus on the case where a rod is perpendicular to the helical axis $\bchi$ and aligned at an angle $\gamma$ away from the $\btau$-axis. This situation is depicted in \fig{numerical_rod_detail}(a) .  Since the rod is perpendicular to the helical axis $\bchi$ we may ignore the effect of chiral twist and parameterize the host director field case within a Cartesian reference frame spanned by the tripod $(\bx, \by, \bz)$ with $\bz = \bchi$:
\beq
\bn_{h}(\bfr) = \bx \cos \Phi(\bfr)  \cos \epsilon(\bfr)+ \by \sin  \Phi(\bfr) \cos \epsilon(\bfr)  + \bz \sin \epsilon(\bfr)
\eeq
 As before,  we ignore end effects and express the spatial variation of the distortion angles in polar coordinates, i.e. $\Phi(r, \vartheta)$ and $\epsilon(r, \vartheta )$ that parameterize space in the lab frame. In principle, the  Euler-Lagrange expressions emerging from minimizing the elastic free energy are strongly coupled and cannot be solved analytically even in the case of weak surface anchoring. We expect, however, that a tilted rod will mostly experience distortions along its main axis $\bchi$, expressed by a non-zero $\epsilon$, while the director deviations $\Phi$ surrounding the lateral cross-section of the rod remain far less affected by the rotation. Then, we can pursue a hybrid route by `constraining $\Phi = \Phi_{0}$ to its solution for the perpendicular case \eq{phioo} and minimize the free energy only with respect to $\epsilon$.
 
 To render the model analytically tractable we assume that the rod cross-section along which the director distortions are expected to occur is curvature-free and can be described by a strip of length $L_{s}$ and $D_{s} \ll L_{s}$. We define a tilt angle $\gamma = \delta - \frac{\pi}{2}$ (with $0<\gamma <\pi/2$) so that $\gamma =0$ corresponds to the case where the rod points perpendicular to the LC host director $\bn$.  All distances are normalized in terms of the colloidal rod diameter $D_{\rm c}$. The distortions are then described by the 2D Laplacian: 
\beq
(\partial_{x}^{2} + \partial_{y}^{2}) \epsilon  = 0
\label{lap2}
\eeq
The general solution reads:
\beq
\epsilon(x,y) = \sum_{n=1}^{\infty} e^{-n \pi x } [ a_{n} \cos(n \pi y) + b_{n} \sin( n \pi y ) ]
\eeq
which vanishes in the far-field limit $\epsilon (x \rightarrow \infty) = 0$. The Rapini-Papoular surface anchoring free energy reads:
\begin{align}
\frac{F_{s}}{KL_{\rm c}} = - \frac{1}{2 \ell_{s}} \int_{0}^{1} dy \cos^{2} (\gamma - \epsilon(0,y))   
\end{align}
which translates into the following boundary condition at the surface of the strip located at $x=0$:
\beq
\partial_{x} \epsilon(0,y) = \frac{1}{4 \ell_{s}} \sin [ 2 (\gamma - \epsilon(0,y)) ]
\eeq
Further, for symmetry reasons we require the distortion angle to be vanishing at both sides of the strip:
\beq
\epsilon(0, 0 ) = \epsilon(0, 1)  =0
\eeq
which implies that $a_{n} =0$. The coefficients $b_{n}$ need to be  resolved from:
\beq
\frac{n \pi}{2} b_{n} = \frac{1}{4 \ell_{s}} \int_{0}^{1} dy \sin (n \pi y) \sin \left [ 2 \left (\gamma - \sum_{k=1}^{\infty} b_{k} \sin (k \pi y) \right ) \right ]  
\label{bcnum}
\eeq
For small tilt angles $\gamma  \ll 1$  distortions are expected to be weak $\epsilon \ll 1$ so that we linearize $\sin 2 (\gamma - \epsilon ) \approx 2( \gamma  - \epsilon)$.
This enables us to resolve the coefficients analytically:
\beq
b_{n} =  \left ( \frac{1 - (-1)^{n}}{(n \pi)^{2}}\right ) \frac{\gamma}{\ell_{s}}  
\eeq
The free energy increase induced by the elastic distortions is given by:
\begin{align}
\Delta F_{el} &= \frac{ \pi KL_{\rm c}}{4} \sum_{n=1}^{\infty} n b_{n}^{2}
\end{align}
which in the linearized regime for small $\gamma$ gives a simple analytical result:
\begin{align}
\Delta F_{el} = \frac{7 KL_{\rm c}}{8 \pi^{3}} \zeta(3) \left ( \frac{\gamma }{\ell_{s}} \right )^{2} 
\label{felgg}
\end{align}
with $\zeta(3) \approx 1.2 $ a constant from the Riemann-Zeta function $\zeta(x)$. The surface anchoring free energy reads:
\begin{align}
F_{s} = - \frac{L_{\rm c}D_{\rm c}W_{0}}{2} 2 \int_{0}^{1} dy \cos^{2} (\gamma - \epsilon(0,y))   
\end{align}
where the factor two reflects the two opposing sides of the rectangular strip with surface $LD$ whose contributions are equivalent. Then, in the absence of elastic distortions and no tilt ($\gamma = 0$) the surface anchoring free energy would simply be $F_{s} = -L_{\rm c}D_{\rm c}W_{0}$ which only marginally differs from the result for the cylindrical case $F_{s} = -(\pi/4)L_{\rm c}D_{\rm c}W_{0}$.
Within the linearized regime for small tilt angles $\gamma \ll 1$  the change in surface anchoring free energy imparted by the elastic distortions is given by:
\begin{align}
\Delta F_{s} &\approx L_{\rm c}D_{\rm c}W_{0}  \int_{0}^{1} dy  (\gamma - \epsilon(0,y))^{2}  \nonumber \\
& \approx W_{0} L_{\rm c}D_{\rm c}  \left ( 1 + \frac{1}{48 \ell_{s}^{2}} - \frac{7  \zeta(3)}{ \pi^{3} \ell_{S}} \right) \gamma^{2} 
\end{align}
This expression along with \eq{felgg} clearly reflects the basic trade-off between surface anchoring and elasticity where the cost in elastic free energy is partly compensated by a reduction of the surface anchoring free energy (last term). The total free energy change for small tilt angles now reads:
\beq
\Delta  F_{\rm tot} \sim  W_{0} L_{\rm c}D_{\rm c} \left ( 1  - \frac{49 \zeta(3)}{8\pi^{3} \ell_{s}}  \right ) \gamma^{2} + \mathcal{O}(\gamma^{2}/\ell_{s}^{2})
\label{ftiltcorr}
\eeq
Let us now compare our results with the simple Rapini-Papoular expression \eq{usurf} in the {\em absence} of elastic distortions. Taking $\theta=\pi/2$ and expanding for small $\gamma $ we find:
 \beq
 \Delta  F_{\rm tot}^{(s)} \sim \frac{\pi}{4} W_{0}L_{\rm c}D_{\rm c} \gamma^{2}
\eeq
Disregarding the trivial curvature prefactor $\pi/4$ in the last expression, we find that the impact of the elastic distortions is rather marginal, since the correction term in  \eq{ftiltcorr} is less than 1 $k_{B}T$. Numerical resolution of \eq{bcnum} reveals that weak elastic distortions  occur mostly when the rod is at an oblique angle $\gamma = \pi/4$.  The predictions from our analytical model are depicted in \fig{numerical_rod_detail}(c).

\section*{Appendix B: Elastic distortions around the disk surface}

Ignoring elastic distortions we find that disks with homeotropic surface anchoring tend to orient along the local molecular director $\bn$, as observed in experiment. This is the optimal situation that incurs the least amount of elastic distortions, compared to the other principal directions in which cases the disk surface would experience strongly unfavorable tangential surface ordering. However, even when the disk normal is aligned along the local nematic director there are local mismatches between the far-field and preferred surface director due to the weak twisting of the host director along the helix axis $\bchi$ and when the rod normal fluctuates away its equilibrium orientation. The elastic distortions are expected to be weak but they will become more outspoken at shorter cholesteric pitches.  It is instructive to compute the extent of these distortions along the lines of our previous analysis for rods. Let us consider an infinitely thin disk with its normal pointing along $\bn$ and rotated over an angle $\delta$ through the helix axis $\bchi$ so that the disk  vector is restricted to lie in the plane perpendicular to it. We assume weak elastic distortions $\Phi$ developing in this plane.  Defining a host director in the Cartesian lab frame $\bn_{h} = \bx \cos \Phi(x,y) + 
\by \sin \Phi(x,y)$ we find, assuming elastic isotropy,  that the distortions are described by the Laplace equation:
\beq
(\partial_{x}^{2} + \partial_{y}^{2}) \Phi  = 0
\label{lap3}
\eeq
The effect of a twisting host director is accounted for through the surface anchoring free energy:
\begin{align}
F_{s} &=  -\frac{W_{0}}{2} \oint d{\mathcal S}  [\bn_{h} \cdot  (\mathcal{R} (qz + \delta) \cdot  \bv({\mathcal S}))]^{2} 
\label{raporotdisc}
\end{align}
where ${\mathcal S} $ parameterizes the face of the disk (as previously we ignore finite thickness effects for disks with $D_{\rm c} \gg L_{\rm c}$) and $\bv = (1,0,0)$ indicating homeotropic anchoring along the surface normal. The rotation matrix reads:
\beq
\mathcal{R}( qz +\delta )  = \begin{pmatrix} 
\cos (qz + \delta)  & -\sin (qz + \delta)  & 0 \\
\sin (qz + \delta) & \cos (qz + \delta) & 0 \\
0 & 0 & 1 \\
\end{pmatrix}
\eeq
A key distinction with the rod case discussed previously is that the distortions are not uniform across the disk surface but depend on the location of the surface element with respect to the helical axis.  It is convenient to divide the disk surface into infinitely thin strips, with each surface element on the strip being equidistant from the centre-of-mass along the helical axis $\bchi$  thus experiencing the same degree of elastic distortions. 

For notational brevity, we implicitly normalize all lengths in units of the disk diameter $D_{\rm c}$ and parameterize the disk surface in terms of $y = \tfrac{1}{2} \cos \alpha$ and $z = \tfrac{1}{2} \sin \alpha$ with $-\pi < \alpha <  \pi$. Each strip then has length $L_{s} = \cos \alpha$ and thickness $D_{s} = \tfrac{1}{2} \cos \alpha d \alpha$ and surface $ds = L_{s}D_{s} $. The surface anchoring free energy of an arbitrary strip with surface $ds$ and centre-of-mass distance $z$ then reads: 
\begin{align}
F_{s}^{\rm strip}  &= -W_{0} [\cos (\Phi(0,y) - qz -\delta )]^{2} ds
\label{sadisc}
\end{align}
The boundary condition at the strip the disk equator ($\alpha=0)$ reads: 
\begin{align} 
\Phi(  \infty, 0 ) & =  0 \nonumber \\
\ell_{s} \partial_{x} \Phi(0, y ) & =   -\frac{1}{2 }   \sin [2 ( \Phi(0,y) - qz - \delta ) ] \nonumber \\ 
& \approx \frac{1}{2} \sin [2(qz + \delta)]  - \cos [2 (qz + \delta)] \Phi(0,y)
\label{bcdisc1}
\end{align}
where we take $0<y<1$ for convenience. The distortions should be symmetric at the edges ($\Phi (0,0)  = \Phi (0, 1)$).  The general solution of the Laplace equation \eq{lap3} reads:
\beq
\Phi(x,y) = \sum_{n=1}^{\infty} e^{-n \pi x }  b_{n} \sin(n \pi y)  
\label{seriesxy}
\eeq
Applying the boundary conditions we obtain the following expression for the coefficients: 
\beq
b_{n}  = \frac{\sin [ 2 (qz+ \delta)  ]}{\cos[2 (qz + \delta)] - n \pi \ell_{s}} \left ( \frac{1-(-1)^{n}}{n \pi}\right ) 
\label{bnscenario1}
\eeq
Given that $q$ and $-q$ do not give equivalent results we conclude that the distortions created near the disk surface carry a distinct chiral signature imparted by the chirality of the host LC, as evidence by the Landau - de Gennes simulations [\fig{chiral_schematic} and \fig{molecular_biaxiality_contour}]. The nature of the imprint depends on the twist angle $\delta $ between the disk normal and the molecular director $\bn$.
We further deduce that the distortions vanish at infinitely weak surface anchoring ($\ell_{s} \rightarrow \infty$) and in the absence of twist and tilting ($q=0$ and $\delta=0$), as we expect. The elastic free energy for the total disk is given by:
\beq
\Delta F_{el} = \frac{\pi KD_{\rm c}}{4}  \int_{-\pi/2}^{\pi/2} d \alpha \cos \alpha \sum_{n} n b^{2}_{n}
\eeq
which may be evaluated as a function of the angle $\delta$ between the disk normal and the molecular director taking the surface anchoring extrapolation length (in units of the disk diameter $D_{\rm c}$) to be about $\ell_{s} \approx 3$.  The change in surface anchoring free energy induced by the distortions follows from linearizing \eq{sadisc} and integrating over all strips:
\begin{align}
\Delta F_{s}  &= \frac{W_{0}D_{\rm c}^{2}}{2} \int_{-\pi/2}^{\pi/2} d \alpha \cos^{2} \alpha \sin[2(qz + \delta)] \nonumber \\ 
&\times \sum_{n} b_{n}  \left ( \frac{1-(-1)^{n}}{n \pi}\right )
\label{sadiscseries}
\end{align}
We reiterate that $z$ depends on the angle $\alpha$ via $z = \tfrac{D_{\rm c}}{2} \sin \alpha$.

We  finish our analysis by considering the case where the disk normal rotates over the $\btau$-axis by an angle $\zeta$. This is equivalent to the situation depicted in \fig{numerical_disk}(c) and (d).  In this situation, the tilting will generate additional weak LC director distortions  across the $\bchi$-direction that we denote by the angle $\epsilon$. The spatially-dependent host director now reads:
\begin{align}
\bn_{h}(\bfr) = & \begin{pmatrix}
\cos \Phi(\bfr)  \cos \epsilon(\bfr)\\
\sin  \Phi(\bfr) \cos \epsilon(\bfr) \\ 
 \sin \epsilon(\bfr)
 \end{pmatrix}
\end{align}
with $\bfr = (x,y) $. Each distortion angle obeys the Laplace equation in the $\bn - \btau$-plane:
\begin{align}
(\partial_{x}^{2} + \partial_{y}^{2}) \Phi  &= 0 \nonumber \\
(\partial_{x}^{2} + \partial_{y}^{2}) \epsilon  &= 0 
\end{align}
The surface anchoring free energy now takes the following form:
\begin{align}
F_{s} &=  -\frac{W_{0}}{2} \oint d{\mathcal S}  [\bn_{h} \cdot  (  \mathcal{R}_{\zeta} \mathcal{R} (qz) \cdot  \bv({\mathcal S}))]^{2} 
\label{raporotdisc2}
\end{align}
where the matrix $\mathcal{R}_{\zeta}$ describes a rotation of the disk normal over the $\btau$-axis (cf. \fig{numerical_disk}(c)):
\beq
\mathcal{R}_{\zeta}= 
\begin{pmatrix} 
\cos \zeta & 0 & \sin \zeta \\
0 & 1 & 0 \\
-\sin \zeta & 0 &  \cos \zeta \\
\end{pmatrix}
\eeq
Analogous to the previous case, we may derive boundary conditions from linearizing $F_{s}$ for weak distortions $\Phi \ll 1$ and $\epsilon \ll1$. Plugging in the general solution [\eq{seriesxy}] and defining $b_{n}$ as the distortion modes pertaining to $\Phi(x,y)$ and $d_{n}$ as those for $\epsilon(x,y)$ we find that both distortion angles are intricately coupled, as expected:
\begin{align}
b_{n} &=  c_{n} \cos \zeta \sin (2 q z) \nonumber \\
d_{n} & = c_{n} \sin (2 \zeta) \cos^{2} (q z) 
\end{align}
From these we immediately assert  the most basic scenarios; both distortions vanish for a disk in an achiral host ($q=0$) at zero tilt ($\zeta=0$), whereas at nonzero tilt angle only $\epsilon(d_{n}) $ is nonzero. For a disk immersed in a chiral host ($q \neq 0 $) at zero tilt ($\zeta =0$) we recover the previous scenario with $\Phi(b_{n})$ given by \eq{bnscenario1} and $\epsilon(d_{n}) =0$]. Both distortion angles are expected to be nonzero in case the disk normal is tilted away from the local director of the chiral host. The common prefactor reads:  
\beq
c_{n} = \frac{ 2\left ( \frac{1-(-1)^{n}}{n \pi}\right )}{1 + 2\ell_{s} n \pi  -\cos( 2 \zeta) - 2 \cos^{2} \zeta \cos (2 q z)}
\eeq
The change in elastic free energy is a simple superposition of amplitudes:
\beq
\Delta F_{el} = \frac{\pi KD_{\rm c}}{4}  \int_{-\pi/2}^{\pi/2} d \alpha \cos \alpha \sum_{n} n \left ( b^{2}_{n} + d_{n}^{2} \right )
\eeq
The contribution arising from the host chirality turns out zero for symmetry reasons:
\begin{align}
 \Delta F_{\rm chiral} &=  K q \int d \bfr \partial_{y} \epsilon (x,y) =0
\end{align}
which is easily inferred from inserting the expansion \eq{seriesxy} and integrating over $y$.  
The reduction in surface anchoring free energy caused by the distortions  $\Phi$ is as follows:
\begin{align}
\Delta F_{s, \Phi} = W_{0}D_{\rm c}^{2} \cos \zeta  \int_{-\pi/2}^{\pi/2} d \alpha \cos^{2} \alpha \sin (2 qz) \nonumber \\
\times \sum_{n} b_{n} \left ( \frac{1-(-1)^{n}}{n \pi}\right )
\end{align}
supplemented with a similar contribution accounting for the distortions $\epsilon$: 
\begin{align}
\Delta F_{s, \epsilon} = W_{0}D_{\rm c}^{2} \sin( 2 \zeta ) \int_{-\pi/2}^{\pi/2} d \alpha \cos^{2} \alpha  \cos^{2} (qz) \nonumber \\
\times \sum_{n} d_{n} \left ( \frac{1-(-1)^{n}}{n \pi}\right )
\end{align}
We find that the surface anchoring is always negative and  outweighs the cost in elastic free energy thus lower the overall free energy of the system, as it should. The results in
\fig{numerical_disk_detail}(c) and (f). We find that the elastic distortions are most developed at oblique orientations $(\delta$ or $\zeta \approx \pi/4)$ and do not strongly depend on the direction along which the disk is rotated.

 If we now reconsider the total alignment potential for disks accounting for corrections derived above we conclude that the ordering of the disks is hardly affected by the distortions. The free energy changes are typically several tens of $k_{B}T$ which is about two orders of magnitude smaller than the typical Rapini-Papoular surface anchoring free energy $ W_{0} D_{\rm c}^{2} $ which is about 1500 $k_{B}T$. disks experiencing weak surface anchoring with  a cholesteric host with large pitch ($qD_{\rm c} <1$)  will therefore simply follow the local molecular director with thermal fluctuations around the optimum angle being strongly suppressed. The considerable penalty incurred  by angular fluctuations away from the local cholesteric director is demonstrated in \fig{numerical_disk_detail}(c) and (f) for a number of different host pitches. Although the presence of  elastic distortions around the disk surface lead to a systematic reduction of the total free energy, their effect on the realigning properties of a colloidal disk immersed in  a cholesteric host LC seems rather marginal.  

 \hide{
 At much short pitches,  comparable to the disk diameter, fluctuations in $\delta$ are greatly facilitated and the local minimum of surface anchoring energy eventually switches over to an equilibrium angle $\delta_{e} = \pi/2$ suggesting  disks  preferentially aligning along the $\btau$-direction, similar to what is observed for homeotropic rods.  Taking the Rapini-Papoular contribution \eq{usp} as the main contribution for the weak surface anchoring regimes considered here we can extract  nematic order parameters associated with such a crossover.  The results, shown in **** \red{We need to decide if and where to show the realignment effect for short-pitch homeotropic disks}, clearly exhibit a sharp transition from uniaxial to biaxial order at a critical pitch of about $qD_{\rm c} \approx 3.8$ which, taking $D_{\rm c} =2 \um$ would correspond to cholesteric pitch of about $p \approx 3.3 \um$.
}

\bibliographystyle{unsrt}
\bibliography{refs}

\end{document}